\documentclass{aastex}
\usepackage{spr-astr-addons}

\def\aap{A\&A\,  }
\def\aj{AJ  }
\def\apj{ApJ\,  }
\def\apjl{ApJ\,  }
\def\apjs{ApJS  }
\def\apss{Astrophysics and Space Science  }

\def\mnras{MNRAS\,  }
\def\pasj{PASJ\,  }
\def\pasp{PASP  }

\RequirePackage{color}

\def\snr{SN \,1993J\,}
\def\sn1987a{SN \,1987A\,}
\def\s1006{SN \,1006\,}
\def\sun{\odot}

\begin{document}

\title
{
On the spherical-axial  transition in
supernova remnants
}
\shorttitle
{
Spherical-axial  transition
}
\shortauthors{Zaninetti}

\author{L. Zaninetti \altaffilmark{1}}
\affil{
Dipartimento di Fisica Generale,   \\
Universit\`a degli Studi di Torino \\
Via Pietro Giuria 1,               \\
           I-10125 Torino, Italy }

\begin{abstract}
A new law of motion for
supernova remnant (SNR) which introduces
the quantity of swept matter  in the thin layer
approximation is introduced.
This new law of motion is tested
on  10 years observations of
\snr.
The introduction of an exponential  gradient
in the surrounding medium  allows to model
an aspherical expansion.
A  weakly asymmetric  SNR, \s1006 ,
and a strongly  asymmetric  SNR, \sn1987a ,
are modeled.
In the case of \sn1987a the   three observed rings
are simulated.
\end{abstract}

\keywords
{
supernovae: general
supernovae: individual (SN 1993J )
supernovae: individual (SN 1006  )
supernovae: individual (SN 1987A )
ISM       : supernova remnants
}

\section{Introduction}
The  theoretical study of  supernova remnant (SNR)
has  been focalized on an expression for the law of motion.
As an example
the Sedov-Taylor  expansion
predicts  $R \propto  t^{0.4}$, see
see \cite{Taylor1950a,Sedov1959,Dalgarno1987}
and the thin layer approximation  in the
presence of a constant density medium
predicts $R \propto  t^{0.25}$,
see \cite{Dyson1997,Dyson1983,Canto2006}.
The very-long-baseline interferometry (VLBI)
observations of \snr
(wavelengths of
3.6, 6, and 18 cm)
show  that
$R\,  \propto  t^{0.82}$  over a 10 year period,
see  \cite{Marcaide2009}.
This  observational fact does  not agree with the
current models  because the radius  of \snr
grows slower than the free expansion
and  faster than the Sedov-Taylor solution,
more details for the spherical case can be
found in  \cite{Zaninetti2011a}.
The SNRs can also be classified at
the light  of the observed symmetry.
A first example is  \snr
which
presented
a
circular symmetry
for 4000
days, see \cite{Marcaide2009}.
An example of weak departure from the circular symmetry
is \s1006  in which a ratio of 1.2
between maximum and minimum radius
has been measured, see \cite {Reynolds1986}.
An example  of axial symmetry  is \sn1987a
in which three rings
are symmetric  in respect to a line
which  connect the centers,
see  \cite{Tziamtzis2011}.
The models cited leave some questions
unanswered or only partially answered:
\begin {itemize}
\item
Is it possible to deduce an equation of motion
for an expanding shell  assuming that
only a fraction of the mass
enclosed in the advancing sphere
is absorbed in the thin layer?
\item
Is it possible to model the complex
three-dimensional (3D)
behavior of
the velocity field of the expanding nebula
introducing an exponential law for the
density ?
\item  Is it possible to make an
       evaluation of the reliability of the
       numerical results
       on radius and velocity compared to the
       observed values?
\item  Can  we reproduce  complicate features
       such as equatorial ring + two outer
       rings  in \sn1987a which are classified
       as a "mystery" ?
\item  Is it  possible to
       build
       cuts of the model intensity which can
       be compared
       with existing
       observations?
\end{itemize}

In order  to answer these questions,
Section \ref{sec_objects}
describes two observed morphologies of SNRs,
Section \ref{sec_classical} reports a new classical
law of motion which  introduces  the concept
of non cubic dependence (NCD) for the mass
included  in the advancing shell,
Section \ref{sec_asymmetry}  introduces
an exponential  behavior  in the number of particles
which  models the aspherical expansion,
Section \ref{sec_application_motion}
applies the  law  of motion
to \sn1987a and  \s1006  introducing the quality of the
simulation,
Section \ref{sec_transfer}
reviews the existing situation
with the radiative transport equation
and Section \ref{sec_image} contains detailed information
on how to build an image of the two astrophysical objects
here considered.

\section{Astrophysical Objects}

\label{sec_objects}

\subsection{A strongly asymmetric SNR , \sn1987a}

The \sn1987a exploded in the Large Magellanic Cloud
in 1987.
The distance of this SN is $\approx~50~kpc$
(163050~$ly$) and a detailed analysis
of the distance , $D$ ,
gives
$D=51.4~kpc$  \cite{Panagia2005}
and
$D=50.18~kpc$  \cite{Mitchell2002}.
In the numerical codes  we will assume
$D=50kpc$ .
The observed image is complex   and we will
follow the  nomenclature of \cite{Racusin2009}
which distinguish  between
torus only , torus +2 lobes and torus + 4 lobes.
In particular we  concentrate  on the torus
which is characterized  by a distance
from the center of the tube
and the radius of the tube.
In Table 2 of \cite{Racusin2009}  is reported the relationship
between distance  of the torus in arcsec and  time
since  the explosion in days.

\subsection{A weakly asymmetric SNR,  \s1006 }

This SN  started to be
visible in 1006 AD
and actually  has a
diameter of 12.7 \mbox{pc}, see \cite{Strom} .
More precisely, on
referring to the radio--map of \s1006  at 1370 \mbox{MHz} by
\cite {Reynolds1986},
it   can be observed that the radius is
greatest in the north--east direction.
From the radio-map
previously mentioned we can  extract the following
observed radii,
$R=6.8~pc$ in the polar  direction and $R= 5.89~pc$ in the
equatorial direction.
 Information on the
thickness of emitting layers is contained in
\cite{Bamba2003}
where the
Chandra observations (i.e., synchrotron X-rays) from \s1006
were analyzed.
The
observations found that  sources of non-thermal radiation are
likely to be thin sheets with a  thickness of about 0.04 pc
upstream  and 0.2 pc downstream of the surface of maximum
emission, which coincide with the locations
of Balmer-line optical
emission , see \cite{Ellison1994} .
The high resolution XMM-Newton Reflection
Grating Spectrometer (RGS)
spectrum of \s1006 gives
two  solutions for the O VII triplet.
One gives  a shell velocity of ~ 6500 $km/s$
  and the second one a shell velocity of 9500 $km/s$,
when a distance of  3.4 kpc is adopted ,see
  \cite{Vink2005}.

\section{Classical law  of motion }
\subsection{Thermal or non thermal emission?}
The synchrotron emission  in SNRs is detected  from $10^8$Hz of
radio-astronomy  to  $10^{19}$Hz of gam\-maastro\-no\-my which means 11
decades in frequency. At the same time  some  particular  effects
such as absorption, transition  from optically thick  to optically
thin medium, line emission,   and energy decay  of radioactive
isotopes ( $^{56}$Ni, $^{56}$Co) can    produce  a change  in the
concavity of the flux  versus frequency relationship, see  the
discussion about Cassiopea A in Section 3.3 of
\cite{Eriksen2009}. A comparison between  non-thermal and thermal
emission (luminosity and surface brightness distribution)  can be
found  in \cite{Petruk2007}, where it is possible to find some
observational tests which allow estimation  of parameters
characterizing the cosmic ray injection on supernova remnant
shocks. At the same time,  a technique  to isolate the synchrotron
from the thermal emission is  widely used, as an example see
X-limb of  SN1006 \cite{Katsuda2010}.

\subsection{The basic assumptions}

The observational dichotomy between  thermal and non thermal
emission influences the use of temperature in the theoretical
models. As an example the Sedov solution is
\begin{equation}
R(t) \approx  0.313\,\sqrt [5]{{\frac {{\it E_{51}}\,{{\it
t_1}}^{2}}{{\it n_0}}} }~{pc}  \quad , \label {rsedov}
\end{equation}
where $t_1$ is the time expressed  in years, $E_{51}$, the  energy
in  $10^{51}$ \mbox{erg} and $n_0$ is   the number density
expressed  in particles~$\mathrm{cm}^{-3}$~ (density~$\rho=n_0$m,
where m = 1.4$m_{\mathrm {H}}$), see
\cite{Sedov1959,Dalgarno1987,Zaninetti2011a}.
 The spectrum of  the emitted
radiation  depends on
 the temperature
behind the shock front , see  for example formula 9.14 in
\cite{mckee},
\begin{equation}
T = \frac{3}{16} \frac{\mu} {k} v_{\mbox{s}}^2 \,\,\, K \quad,
\end{equation}
where $\mu$ is the mean mass per particle, {\it k}
 the Boltzmann constant
and $v_{\mbox{s}}$ the shock velocity expressed in
$\mbox{cm~sec}^{-1}$. On identifying  $v_s$ with the velocity
derived from Eq. \ref{rsedov}
 the Sedov solution
 has the following behavior in temperature
\begin{equation}
T=\frac{3.41\,10^{11}{{\it E_{51}}}^{2/5}} {{{\it n_0}}^{2/5}{{\it
t_1}}^{6/5}}\,\,\,K \quad  ,
\end{equation}
when $\mu$ is the mass of the hydrogen.
 The Sedov approach cannot
be applied to the aspherical case because we do no know the
complex volume occupied during the expansion. A numerical approach
developed to follow the evolution of the superbubbles first
compute the volume occupied during the expansion and then the
pressure , see \cite{Zaninetti2004}. Conversely the thin layer
approximation , see \cite{Dyson1997,Dyson1983},
 takes in consideration
only the swept mass and the velocity is deduced applying the
momentum conservation . Due to the absence of the temperature and
pressure the thin layer approximation  can be classified as a non
thermal model.
\subsection{The incomplete thin layer approximation}
\label{sec_classical}
The thin layer approximation
with non cubic dependence (NCD) , $p$ ,
in classical physics
assumes that only a fraction
of the total mass enclosed in the  volume
of the expansion
accumulates  in a thin shell just after
the shock front.
The  global mass between  $0$ and $R_0$ is    $\frac {4}{3} \pi
\rho  R_0^3  $ where $\rho$ is the density of the ambient medium.
The swept mass included in the  thin layer which characterizes the
expansion is
\begin{equation}
M_0 =( \frac {4}{3} \pi \rho  R_0^3)^{\frac{1}{p}}
\quad  .
\end{equation}
The  mass swept between  $0$ and $R$
is
\begin{equation}
M =( \frac {4}{3} \pi \rho  R^3)^{\frac{1}{p}}
\quad  .
\end{equation}
The conservation of radial momentum requires that,
after the initial  radius $R_0$,
\begin{equation}
M   V =
M_0 V_0
\quad ,
\end{equation}
where $R$ and $V$   are  the
radius and velocity
of the advancing shock.
In classical physics,
the velocity as a function of radius
is:
\begin{equation}
\label{velocityclassical}
V=  V_0 (\frac {R_0}{R})^{\frac{3}{p}}
\quad ,
\end{equation}
and  introducing
$\beta_0= \frac {V_0}{c} $
and
$\beta= \frac {V}{c} $
we obtain
\begin{equation}
\beta =  \beta_0 (\frac {R_0}{R})^{\frac{3}{p}}
\quad .
\label{eqnbeta}
\end{equation}
The law of motion is:
\begin{eqnarray}
R(t) = \nonumber  \\
\left( {R_{{0}}}^{1+3\,{p}^{-1}}+ \left( 3+p \right) v_{{0}}{R_{{0}}}
^{3\,{p}^{-1}} \left( t-t_{{0}} \right) {p}^{-1} \right) ^{{\frac {p}{
3+p}}}
\quad ,
\label{rtclassical}
\end{eqnarray}
where $t$ is time and $t_0$ is
the initial  time.

is:
\begin{eqnarray}
V  =  \frac{N}{D}                                 \\
where                             \nonumber  \\
N=
                                  \nonumber  \\
( {{\it R_0}}^{1+3\,{p}^{-1}}+ ( 3+p  ) \times  \nonumber \\
\times  {\it V_0}\,{{
\it R_0}}^{3\,{p}^{-1}} ( t-{\it t_0}  ) {p}^{-1}  ) ^{{
\frac {p}{3+p}}}{\it V_0}\,{{\it R_0}}^{3\,{p}^{-1}}p
\nonumber \\
D={{\it R_0}}^{1+3\,{p}^{-1}}p+3\,{\it V_0}\,{{\it R_0}}^{3\,{p}^{-1}}t+{
\it V_0}\,{{\it R_0}}^{3\,{p}^{-1}}tp
\nonumber  \\
-3\,{\it V_0}\,{{\it R_0}}^{3\,{p}^{-
1}}{\it t_0}-{\it V_0}\,{{\it R_0}}^{3\,{p}^{-1}}{\it t_0}\,p
\, .
                        \nonumber \\
\label{velocitym}
\nonumber
\end{eqnarray}
Equation (\ref{rtclassical}) can  also be solved
with a similar solution of type
$R=K(t-t_0)^{\alpha}$,
$k$ being a constant,
and the classical result is:
\begin{equation}
\label{radiussimilar}
R(t) =
 \left( \frac{  \left( 3+p \right) V_{{0}}{R_{{0}}}^{3/p}
\left( t-t_
{{0}} \right) }{p}  \right) ^{{\frac {p}{3+p}}}
\quad .
\end{equation}
The similar  solution  for the  velocity is
\begin {eqnarray}
V(t) =       \nonumber  \\
 \left( 3+p \right) ^{-\frac {3} { 3+p}}{p}
^{\frac {3}{3+p}}
{{\it V_0}}^{{\frac {p}{3+p}}}{{\it R_0}}
^{\frac{3}{3+p}}
\left( t-{\it t_0} \right) ^{-\frac {3} { 3+p}}
\quad  .
\label{velocitysimilar}
\end{eqnarray}

\begin{figure*}
\includegraphics[width=6cm]{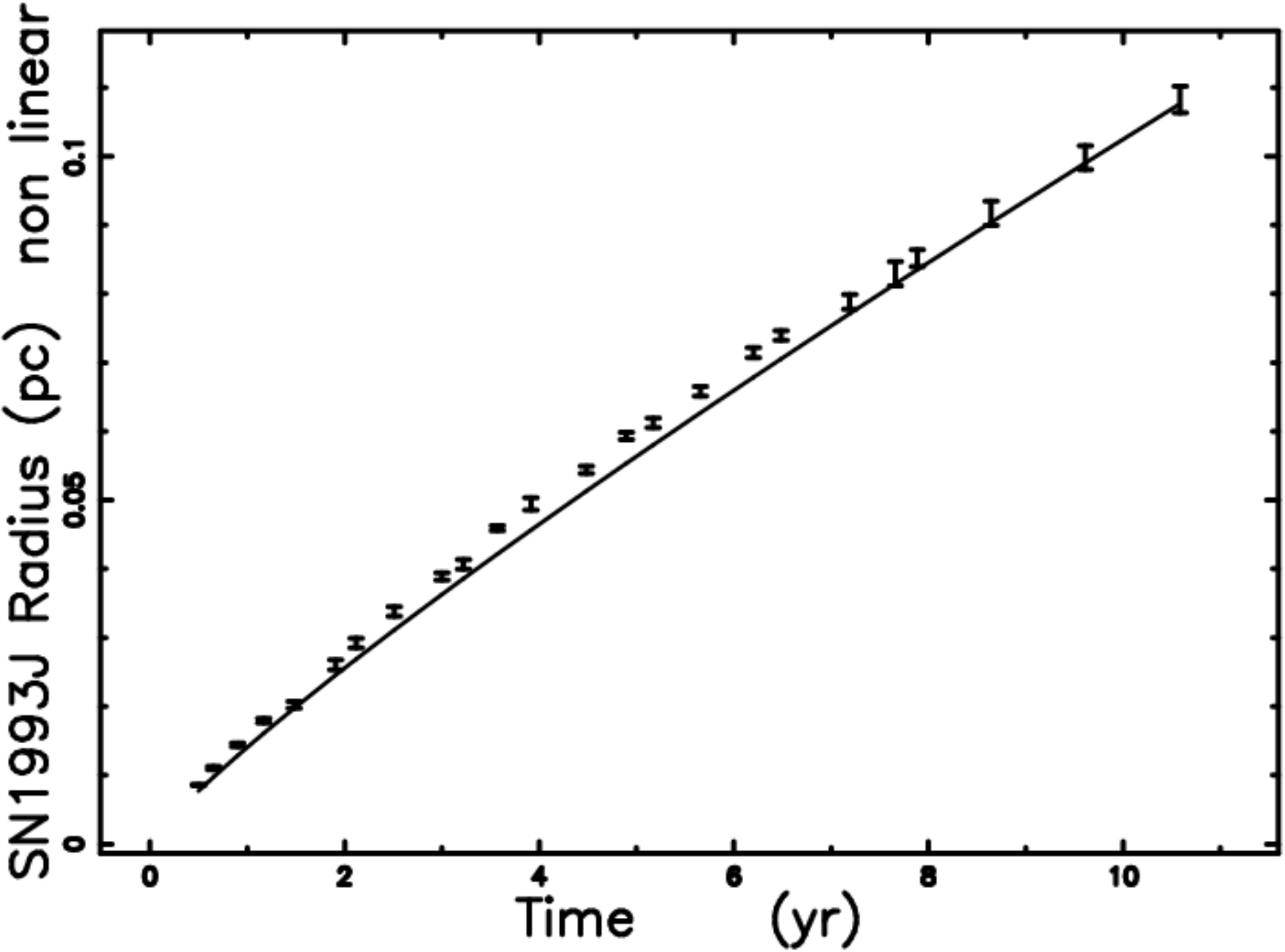}
\caption
{
Theoretical radius as obtained
by the solution of  the
equation (\ref{rtclassical})
(full line), data  and
merit function , $\chi^2$ ,
as in   Table \ref{datafitsn1993j}.
The astronomical data of \snr are represented through
empty stars.
}
\label{fit_lev_1993j}
    \end{figure*}

The similar formula (\ref{radiussimilar})
for the radius  can be compared
with  the observed radius
radius-time  relationship of the  Supernova
reported as
\begin{equation}
R(t) = r_{obs} t^{\alpha_{obs}}
\label{rpower}
\quad ,
\end{equation}
where the two parameters $r_{obs}$
and $\alpha_{obs}$ are
found from the numerical analysis
of the observational
data.
In this case the velocity  is
\begin{equation}
V(t) = r_{obs} \, \alpha_{obs} t^{(\alpha_{obs}-1)}
\quad .
\label {vpower}
\end{equation}
The comparison between theory and astronomical observations allows
to deduce  $p$ , the  NCD parameter ,
 as
\begin{equation}
p  = \frac{3 \alpha_{obs}}{ 1 -\alpha_{obs}} \quad   ,
\end{equation}
where $\alpha_{obs}$ characterizes  the radius-time relationship
in SNRs as given by Eq. (\ref{rpower}). 
The Supernova  \snr represents a test
for this  theoretical
model  of the expansion.
A careful analysis  of the radius-time relationship
for  \snr  shows that
$R(t) \approx 0.015 t^{0.82} \, pc$
when the time is  expressed in $yr$ .
A first  possible system of units
which allows to make a comparison
with the observations is represented by
    $pc$ for the length
and $yr$   for the time.
The theoretical solution
as given  by equation (\ref{rtclassical})
can be found through
the  Levenberg--Marquardt  method ( subroutine
MRQMIN in \cite{press})
and  Fig. \ref{fit_lev_1993j} reports
a numerical  example.k

\begin{table}
\caption {
Numerical value of the parameters
of the fits for \snr and $\chi^2$.
$N$  represents the number of  free parameters.
PL stands for power law  and  NLR  for  non linear radius.
 }
 \label{datafitsn1993j}
 \[
 \begin{array}{cccc}
 \hline
 \hline
 \noalign{\smallskip}
keyword    &  N
& values  & \chi^2         \\
 \noalign{\smallskip}
 \hline
 \noalign{\smallskip}
PL      & 2   &
\alpha_{obs} = 0.82 ;
r_{obs} = 0.015~pc  &  6364 \\
 \noalign{\smallskip}
 \hline
NLR     & 4   &
p=18.8  ;
& 623.5   \\
~       & ~   &
r_{0} = 2.25\,10^{-8}~pc ;
& ~     \\
\noalign{\smallskip}
~     & ~   &
t_{0}=4.98\,0^{-8} ~yr     ;
& ~     \\
~     & ~   &
v_0 =100 000 \frac{km}{s}
& ~     \\
\noalign{\smallskip}
 \hline
 \hline
 \end{array}
 \]
 \end {table}
The quality of the fits is measured by the
merit function
$\chi^2$
\begin{equation}
\chi^2  =
\sum_j \frac {(R_{th} -R_{obs})^2}
             {\sigma_{obs}^2}
\quad ,
\label{chisquare}
\end{equation}
where  $R_{th}$, $R_{obs}$ and $\sigma_{obs}$
are the theoretical radius, the observed radius and
the observed uncertainty respectively.

The conservation of classical  momentum
here adop\-ted
does not take
into account the momentum carried away by photons
or in other words the radiative losses are included
in the NCD  exponent.

\section{Asymmetrical  law  of motion with NCD}

\label{sec_asymmetry}
Given the Cartesian   coordinate system
$(x,y,z)$ ,
the plane $z=0$ will be called equatorial plane
and in  polar coordinates $z= R \sin ( \theta) $,
 where
$\theta$ is the polar angle
and $R$ the distance from
the origin .
The presence of a non homogeneous medium in which
the expansion
takes place is here  modeled
assuming an exponential
behavior for the number of particles of the type
\begin{equation}
n (z) = n_0 \exp {- \frac {z}{h} }
\quad
= n_0 \exp {- \frac {R\times \sin (\theta) }{h} }
\quad  ,
\end{equation}
where  $R$ is the radius of the shell,
$n_0$ is the number
of particles at $R=R_0$ and $h$ the scale.
The 3D expansion will be characterized by the following
properties
\begin {itemize}
\item Dependence of the momentary radius of the shell
      on  the polar angle $\theta$ that has a range
      $[-90 ^{\circ}  \leftrightarrow  +90 ^{\circ} ]$.

\item Independence of the momentary radius of the shell
      from  $\phi$ , the azimuthal  angle  in the x-y  plane,
      that has a range
      $[0 ^{\circ}  \leftrightarrow  360 ^{\circ} ]$.
\end {itemize}
The mass swept, $M$,  along the solid angle
$ \Delta\;\Omega $,  between 0 and $R$ is
\begin{equation}
M(R)=
\frac { \Delta\;\Omega } {3}  m_H n_0 I_m(R)
+ \frac{4}{3} \pi R_0^3 n_0 m_H
\quad  ,
\end {equation}
where
\begin{equation}
I_m(R)  = \int_{R_0} ^R r^2 \exp { - \frac {r \sin (\theta) }{ h}  } dr
\quad ,
\end{equation}
where $R_0$ is the initial radius
and $m_H$ the mass of the hydrogen.
The integral is
\begin{eqnarray}
I_m(R)  =  \nonumber \\
\frac
{
h \left( 2 {h}^{2}+2 R_0h\sin \left( \theta \right) +{R_0}^{2} \left(
\sin \left( \theta \right)  \right) ^{2} \right) {{\rm e}^{-{\frac {R_0
\sin \left( \theta \right) }{h}}}}
}
{
\left( \sin \left( \theta \right)  \right) ^{3}
}
           \nonumber\\
- \frac
{
h \left( 2 {h}^{2}+2 Rh\sin \left( \theta \right) +{R}^{2} \left(
\sin \left( \theta \right)  \right) ^{2} \right) {{\rm e}^{-{\frac {R
\sin \left( \theta \right) }{h}}}}
}
{
\left( \sin \left( \theta \right)  \right) ^{3}
}
\quad .
\end{eqnarray}
The conservation of the momentum gives
\begin{equation}
(M(R))^{\frac{1}{p}}    \dot {R}=
(M(R_0))^{\frac{1}{p}}  \dot {R_0}
\quad  ,
\end{equation}
where $\dot {R}$  is the  velocity
at $R$ and
$\dot {R_0}$  is the  initial velocity at $R=R_0$
where  $p$   is the NCD parameter.
This  means  that only a fraction
of the total mass enclosed in the  volume
of the expansion
accumulates in a thin shell just after
the shock front.
According to the previous expression
the velocity is
\begin{eqnarray}
V(R) =
{{\it R_0}}^{3 {p}^{-1}}{\it V_0}  (  ( -2 {h}^{3}{{\rm e}^
{-{\frac {R\sin ( \theta ) }{h}}}}
\nonumber \\
-2 {h}^{2}{{\rm e}^{-{
\frac {R\sin ( \theta ) }{h}}}}R\sin ( \theta )
\nonumber  \\
-h{{\rm e}^{-{\frac {R\sin ( \theta ) }{h}}}}{R}^{2}
 ( \sin ( \theta )  ) ^{2}+2 {h}^{3}{{\rm e}^{-
{\frac {{\it R_0} \sin ( \theta ) }{h}}}}
\nonumber\\
+2 {h}^{2}{
{\rm e}^{-{\frac {{\it R_0} \sin ( \theta ) }{h}}}}{\it R_0}
 \sin ( \theta ) +h{{\rm e}^{-{\frac {{\it R_0} \sin
 ( \theta ) }{h}}}}{{\it R_0}}^{2} ( \sin ( \theta
 )  ) ^{2}
\nonumber\\
+{{\it R_0}}^{3} ( \sin ( \theta
 )  ) ^{3} )  ( \sin ( \theta )
 ) ^{-3} ) ^{-{p}^{-1}}
\quad  .
\end{eqnarray}
In this differential equation of the first order in $R$ the
variable can be separated and the integration
term by term gives
\begin{equation}
\int_{R_0}^{R} ( M(r))^{1/p}  dr =
(M(R_0))^{1/p} \dot {R_0} \times ( t-t_0)
\quad  ,
\end{equation}
where  $t$ is the time and $t_0$ the time at $R_0$.
The resulting non linear equation ${\mathcal{F}}_{NL}$
expressed in astrophysical units
is
\begin{eqnarray}
{\mathcal{F}}_{NL} =
\int_{R_{0,pc}}^{R_{pc}}
{3}^{-{p}^{-1}} (  ( -2 {h_{pc}}^{3}{{\rm e}^{-{\frac {{r_{pc}}\sin
 ( \theta ) }{h_{pc}}}}}
\nonumber\\
-2 {h_{pc}}^{2}{{\rm e}^{-{\frac {{r_{pc}}\sin
 ( \theta ) }{h_{pc}}}}}{r_{pc}}\sin ( \theta )
\nonumber \\
-h{{\rm e}^{
-{\frac {{r_{pc}}\sin ( \theta ) }{h_{pc}}}}}{{r_{pc}}}^{2} ( \sin
 ( \theta )  ) ^{2}
\nonumber\\
+2 {h_{pc}}^{3}{{\rm e}^{-{\frac {{
\it {R_{0,pc}}} \sin ( \theta ) }{h_{pc}}}}}+2 {h_{pc}}^{2}{{\rm e}^{-{
\frac {{\it {R_{0,pc}}} \sin ( \theta ) }{h_{pc}}}}}{\it {R_{0,pc}}} \sin
 ( \theta )
\nonumber\\
 +h{{\rm e}^{-{\frac {{\it {R_{0,pc}}} \sin (
\theta ) }{h_{pc}}}}}{{\it {R_{0,pc}}}}^{2} ( \sin ( \theta
 )  ) ^{2}
\nonumber  \\
+{{\it {R_{0,pc}}}}^{3} ( \sin ( \theta
 )  ) ^{3} )  ( \sin ( \theta )
 ) ^{-3} ) ^{{p}^{-1}}  dr
\nonumber\\
- 1.02 10^{-5} {3}^{-{p}^{-1}}
{{\it R_{0,pc}}}^{3 {p}^{-1}}{\it \dot {R}_{0,kms}}
 ( t_1-{\it t_{0,1}} ) =0
\label{nonlinearastro}
\end{eqnarray}
where   $t_1$  and
$t_{0,1}$ are  $t$ and $t_0$
expressed  in  \mbox{yr} units,
$r_{pc}$ and  $R_{0,pc}$  are
$R$ and  $R_0$  expressed in  $pc$,
$\dot {R}_{0,kms}$
is
  $\dot{R}_0$   expressed
in $\frac{km}{s}$,
$\theta$ is expressed in radians
and  $h_{pc}$ is  the  the scale , $h$  ,
expressed in $pc$.
It is not possible to find  $R_{pc}$   analytically  and
a numerical method   should be implemented.
In  our case in order
to find  the root of  ${\mathcal{F}}_{NL}$,
the FORTRAN SUBROUTINE  ZRIDDR from \cite{press} has been used.

\section{Applications of the law of motion}

\label{sec_application_motion}
From a practical point  of view, $\epsilon$ ,
the percentage  of
reliability  of our code is   introduced,
\begin{equation}
\epsilon  =(1- \frac{\vert( R_{\mathrm {pc,obs}}- R_{pc,\mathrm {num}}) \vert}
{R_{pc,\mathrm {obs}}}) \cdot 100
 ,
\label{efficiency}
\end{equation}
where $R_{pc,\mathrm {obs}}$ is the   radius as given
by the astronomical observations in parsec ,
and  $R_{pc,\mathrm {num}}$ the radius
obtained from our  simulation
in parsec.

\subsection{Results on the strongly asymmetric  \sn1987a}

\begin{figure}
\includegraphics[width=6cm]{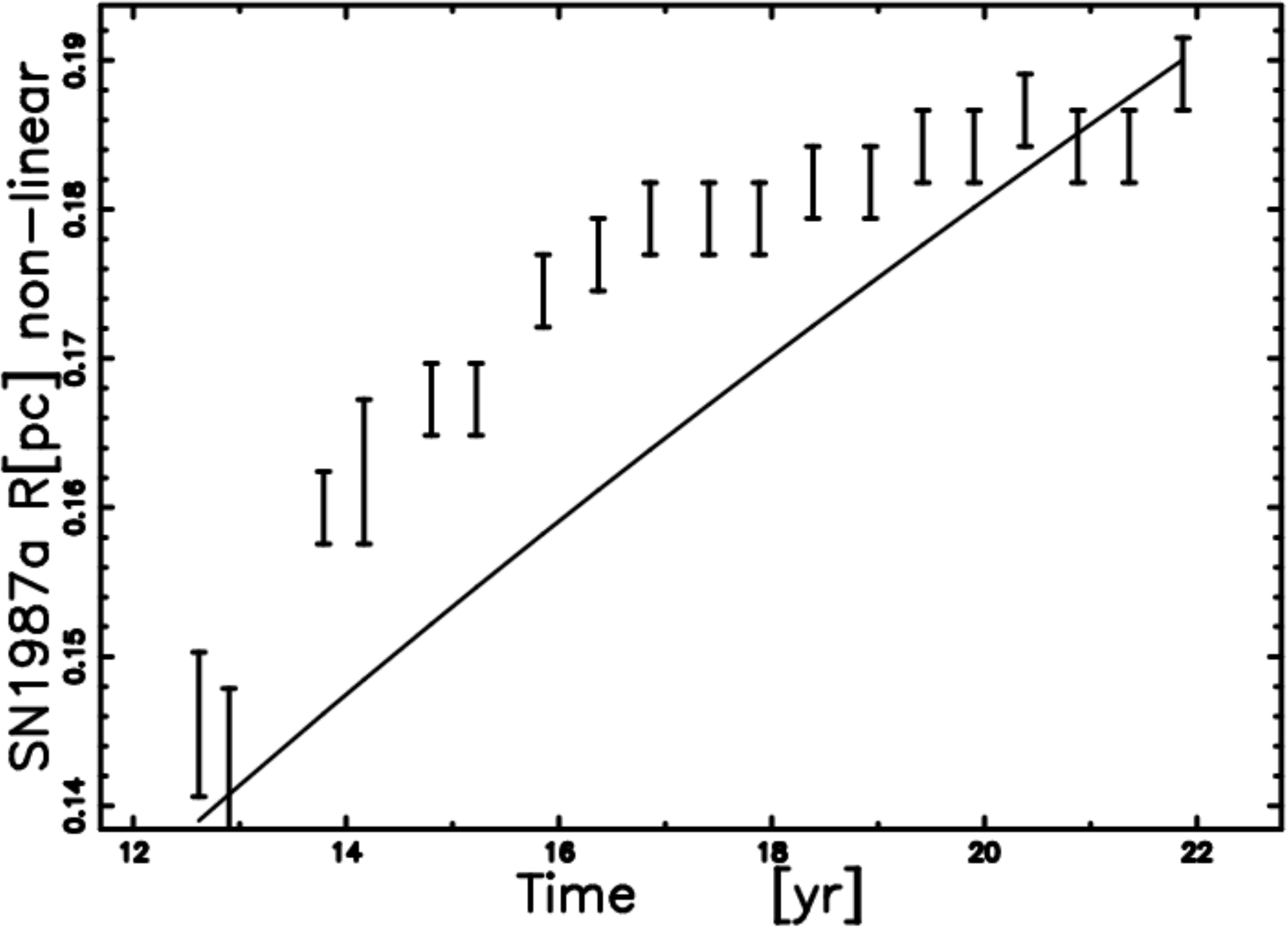}
\caption {
Radius as a function of time for an
exponentially varying medium (full line)
when $\theta=0.001$ (equatorial plane)
and  astronomical data of torus only  as  extracted
from  Table 2 in  \cite{Racusin2009}.
Physical parameters as in Table \ref{datafit1987a}.
          }%
    \label{radiusptheta1987a}
    \end{figure}

\begin{figure}
\includegraphics[width=6cm]{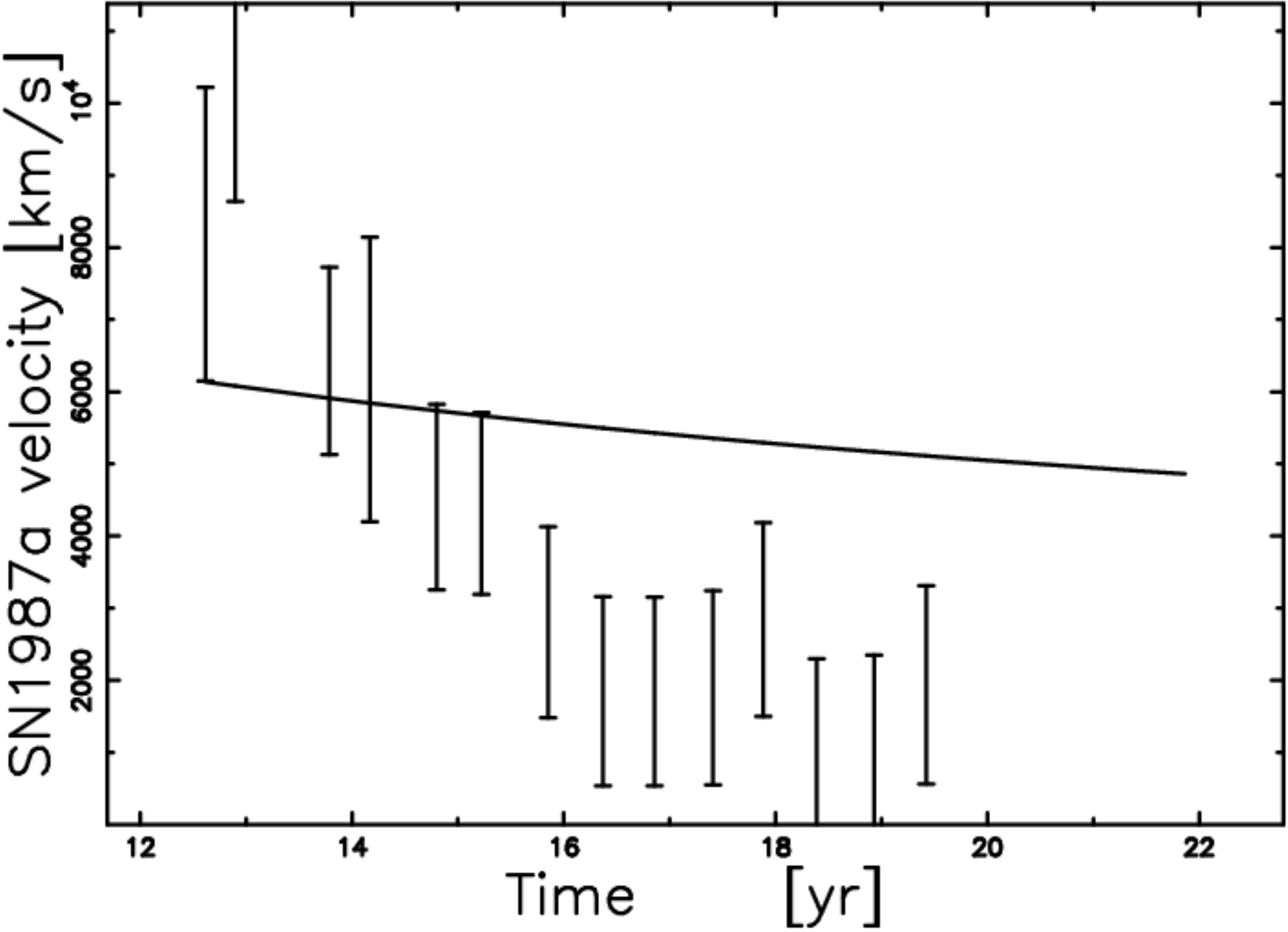}
\caption {
Velocity  as a function of the time for an
exponentially varying medium (full line)
when $\theta=0.001$ (equatorial plane)
and  astronomical
velocities as deduced
from  Table 2 in  \cite{Racusin2009}.
Physical parameters as in Table \ref{datafit1987a}.
          }%
    \label{veltheta_fit1987a}
    \end{figure}

The first target  is to simulate  the torus only
of  \sn1987a
(our equatorial  plane)
and  this operation allows  to calibrate our code ;
Table \ref{datafit1987a}  reports the input data,
Figs.  \ref{radiusptheta1987a} and
\ref{veltheta_fit1987a} the  behavior of
radius and velocity  respectively  as function
of the time.
Due to the  complexity  of the 3D  structure
we  report  the reliability  of the simulation
only in the equatorial  plane  (torus  only)
which  is  $\epsilon=91.4 \%$ for the radii
and  $\epsilon=38.3 \%$ for the velocity.
\begin{table}
\caption
{
Numerical value of the parameters
of the simulation  for  \sn1987a
}
\label{datafit1987a}
 \[
 \begin{array}{lc}
 \hline
 \hline
 R_{0,pc}         &   0.014   \\
\dot {R}_{0,kms}  &   26000  \\
p                 &   4      \\
t_{0,1}           &   0.0218 \\
t_{1}             &   21.86  \\
\noalign{\smallskip}
 \hline
 \hline
 \end{array}
 \]
 \end {table}

\begin{figure}
\includegraphics[width=6cm]{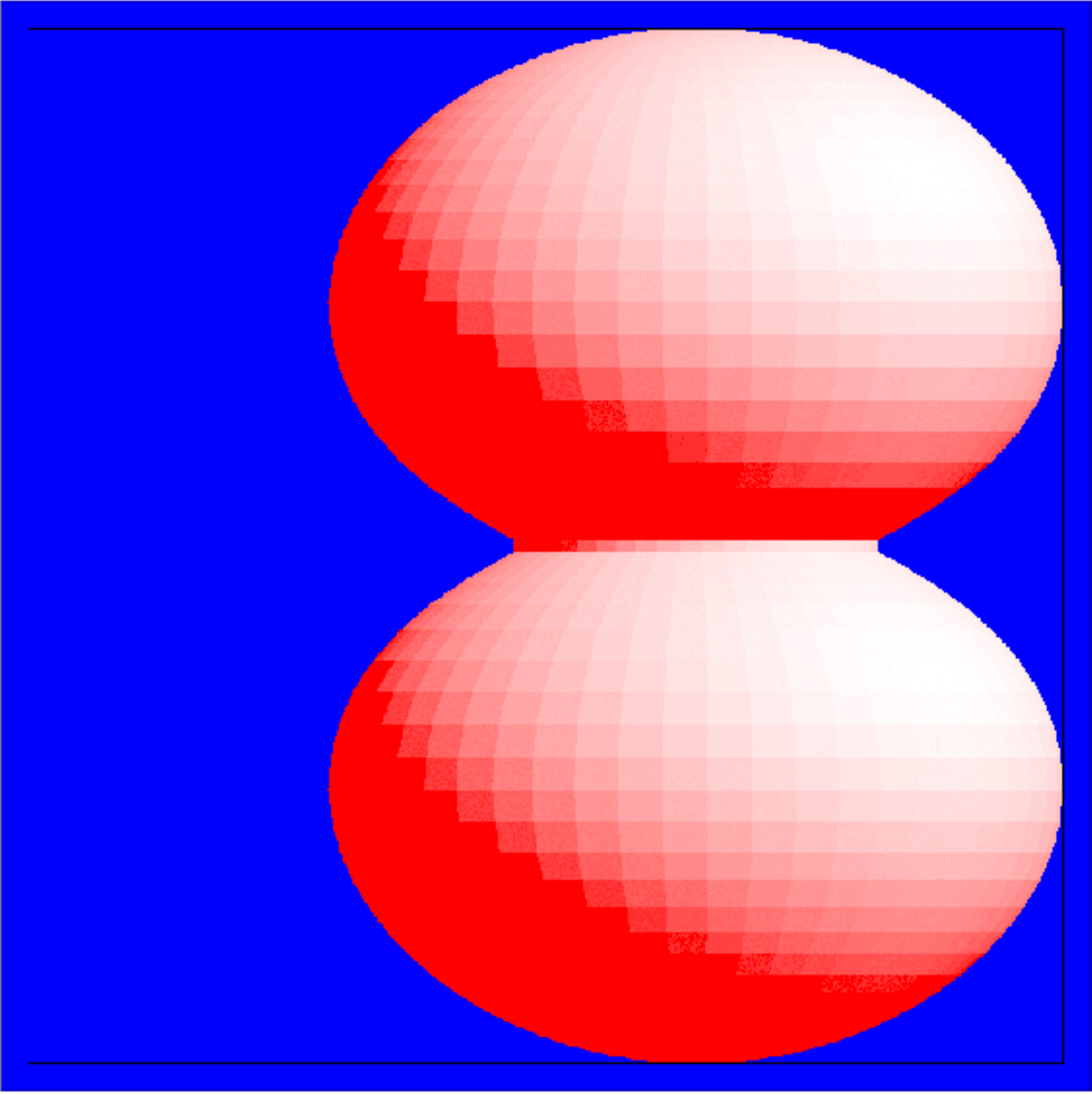}
\caption { Continuous  three-dimensional surface of \sn1987a : the
three Eulerian angles characterizing the point of view are
     $ \Phi   $=90    $^{\circ }  $,
     $ \Theta $=90    $^{\circ }  $
and  $ \Psi   $=90    $^{\circ }  $. Physical parameters as in
Table \ref{datafit1987a}.
          }%
    \label{1987a_faces}
    \end{figure}

\begin{figure}
\includegraphics[width=6cm]{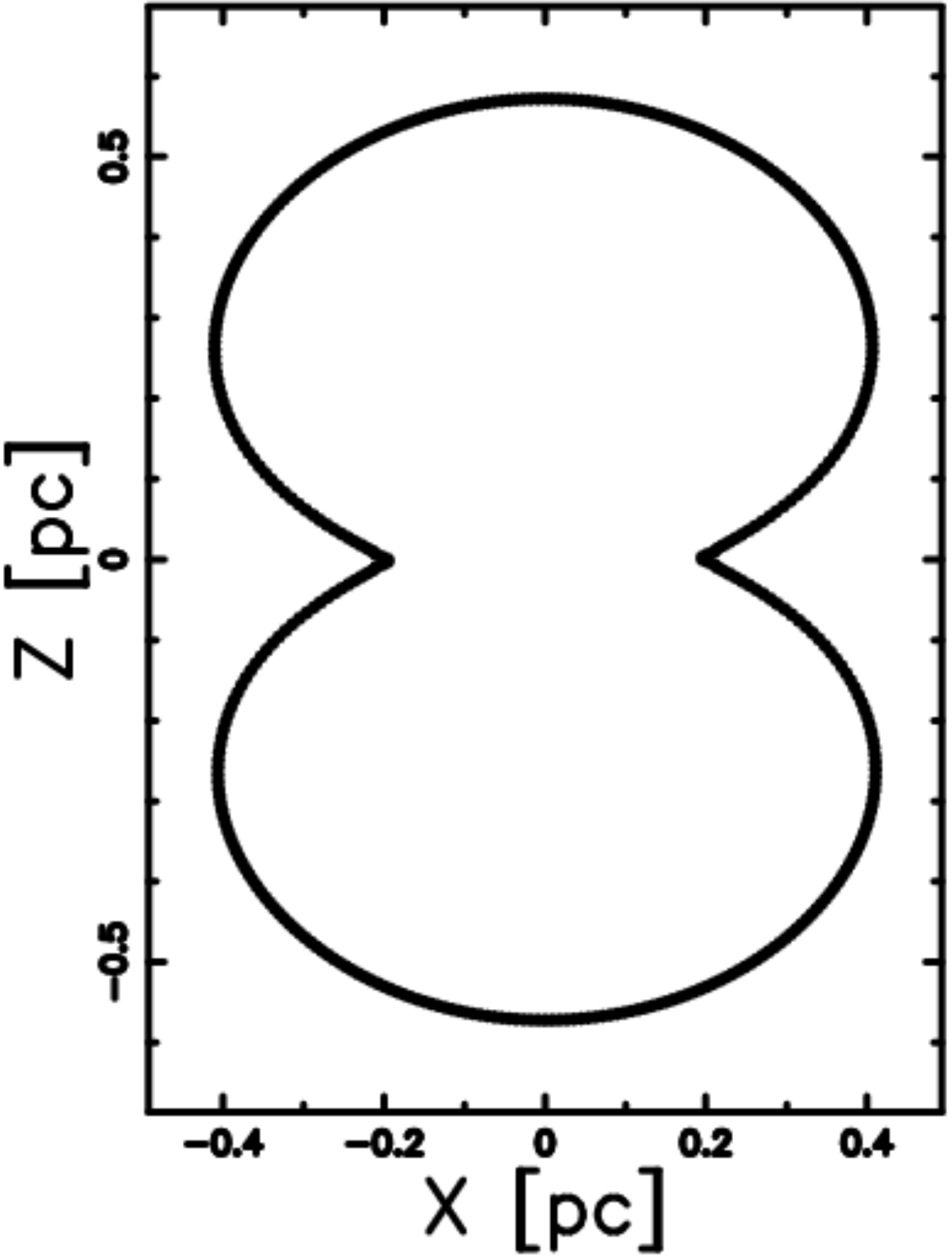}
\caption { Section of  \sn1987a  in  the
{\it x-z}  plane.
The horizontal and vertical axis are in $pc$.
Physical parameters
as in Table \ref{datafit1987a}.
          }%
    \label{1987a_cut}
    \end{figure}

\begin{figure}
\includegraphics[width=6cm]{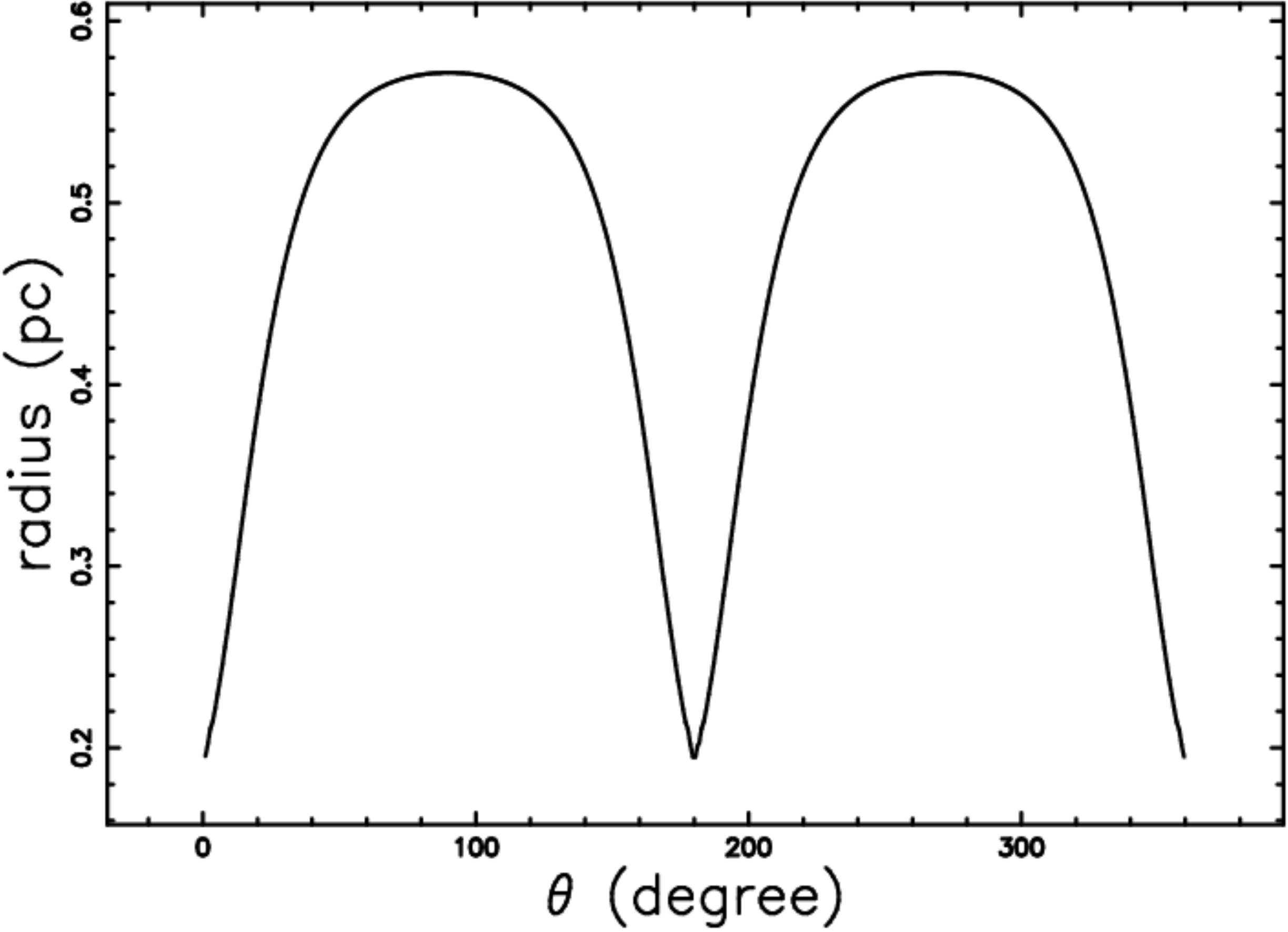}
\caption {
Radius in $pc$  of \sn1987a as a function of
the position angle in degrees.
Physical parameters as in
Table \ref{datafit1987a}.
          }%
    \label{1987a_radius}
    \end{figure}

\begin{figure}
\includegraphics[width=6cm]{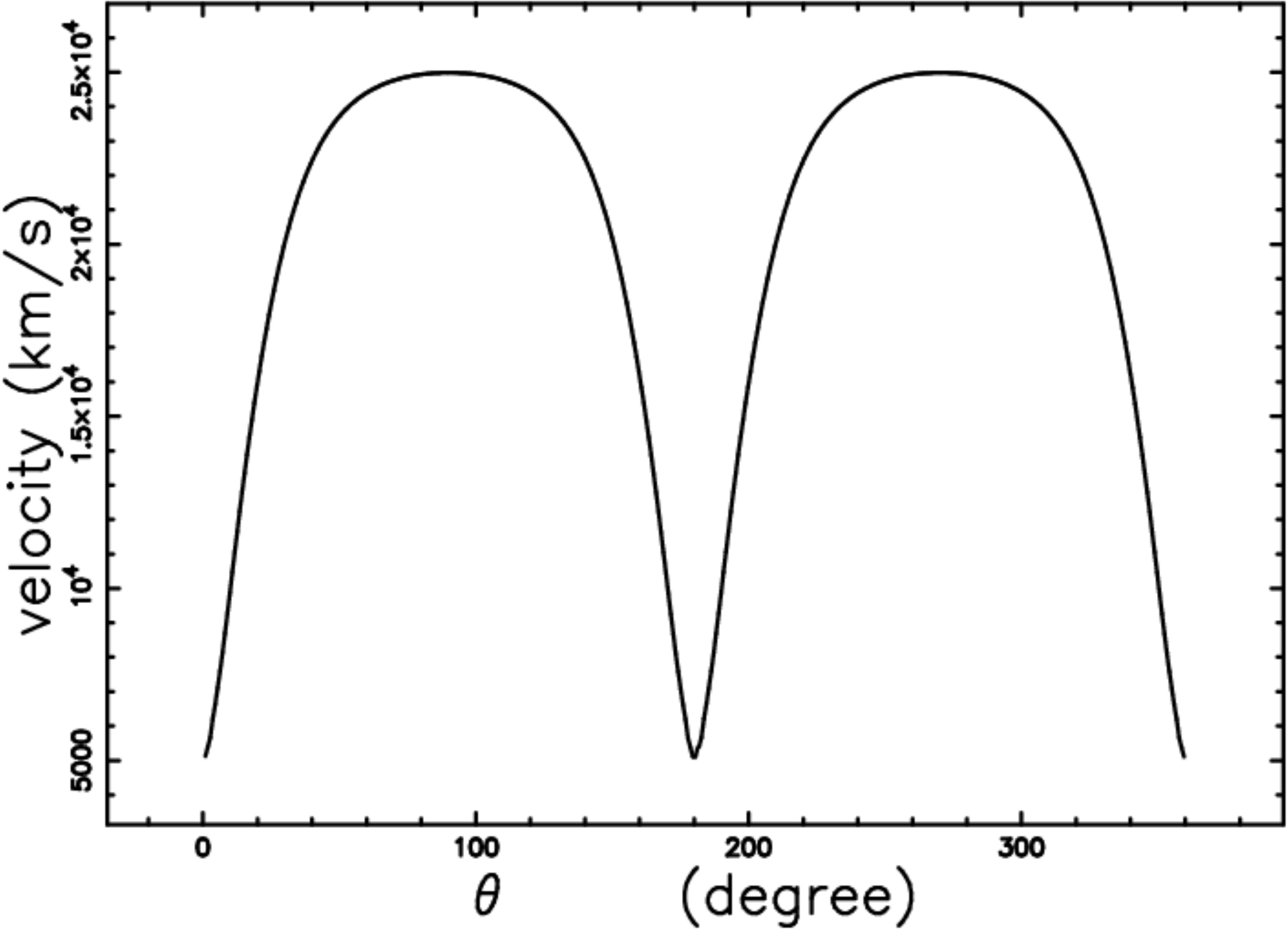}
\caption {
Velocity  in $\frac{km}{s}$  of \sn1987a as a
function of the position angle in degrees.
Physical parameters as
in Table \ref{datafit1987a}.
          }%
    \label{1987a_velocity}
    \end{figure}

\begin{figure}
\includegraphics[width=6cm]{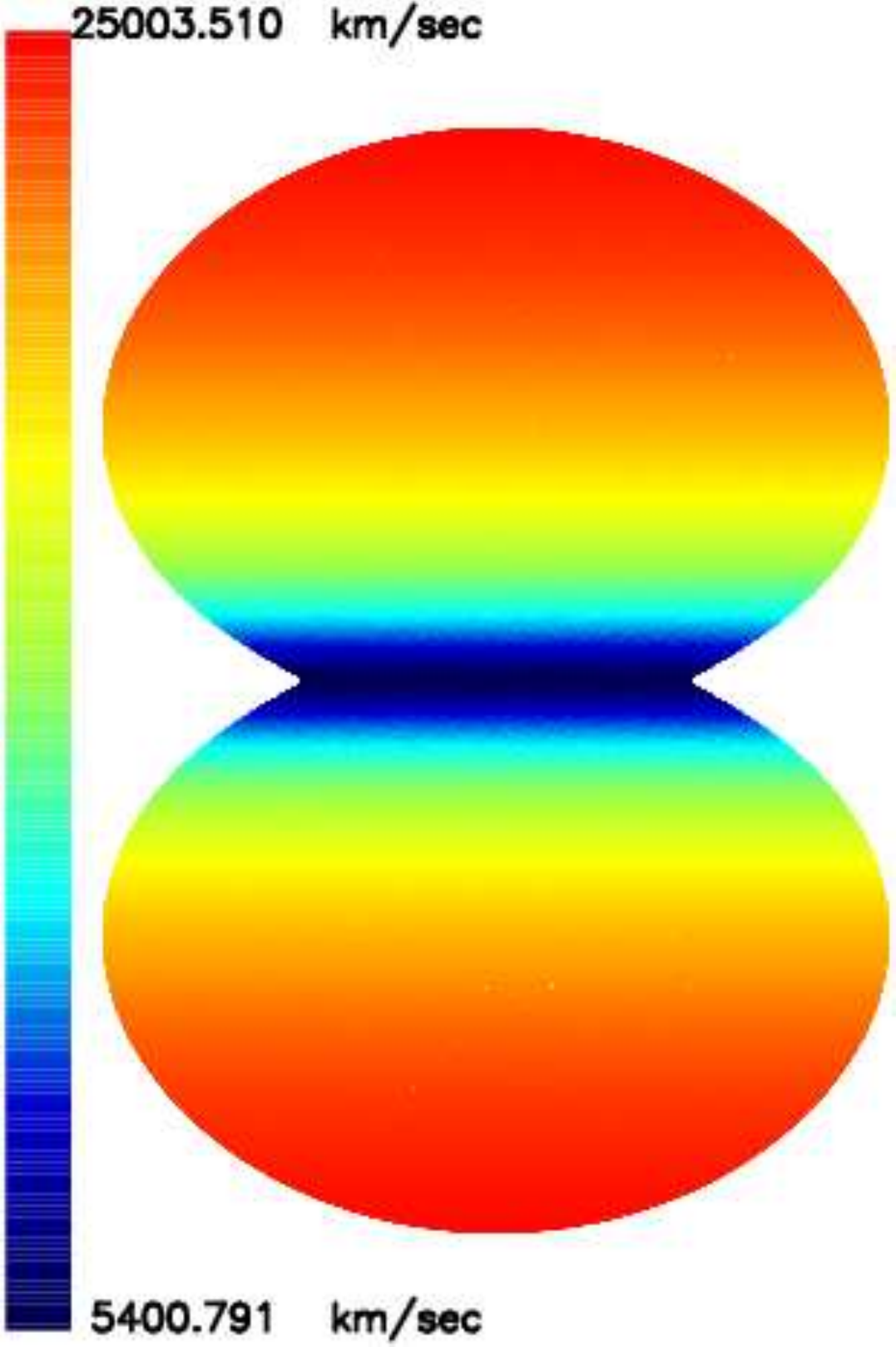}
\caption {
  Map of the expansion velocity
  in $\frac{km}{s}$
  relative to the simulation of \sn1987a
  when 300000 random points are
  selected on the surface.
  Physical parameters as in Table \ref{1987a_velocity}.
          }%
    \label{1987a_velocity_field}
    \end{figure}

After  this calibration on the  equatorial plane
we continue identifying the lobes
of  \sn1987a as bipolar SNR as seen
from a given   point of view.
The complex 3D behavior of the advancing SNR
is reported
in Fig. \ref{1987a_faces}
and Fig. \ref{1987a_cut}
reports the asymmetric expansion in
a  section crossing the center.
In order to better visualize the asymmetries
Figs. \ref{1987a_radius} and
 \ref{1987a_velocity}
report the radius  and the velocity  respectively
as a function of the position angle $\theta$.
The combined effect of spatial asymmetry and field of velocity are
reported in Fig. \ref{1987a_velocity_field}.

\begin{figure}
\includegraphics[width=6cm]{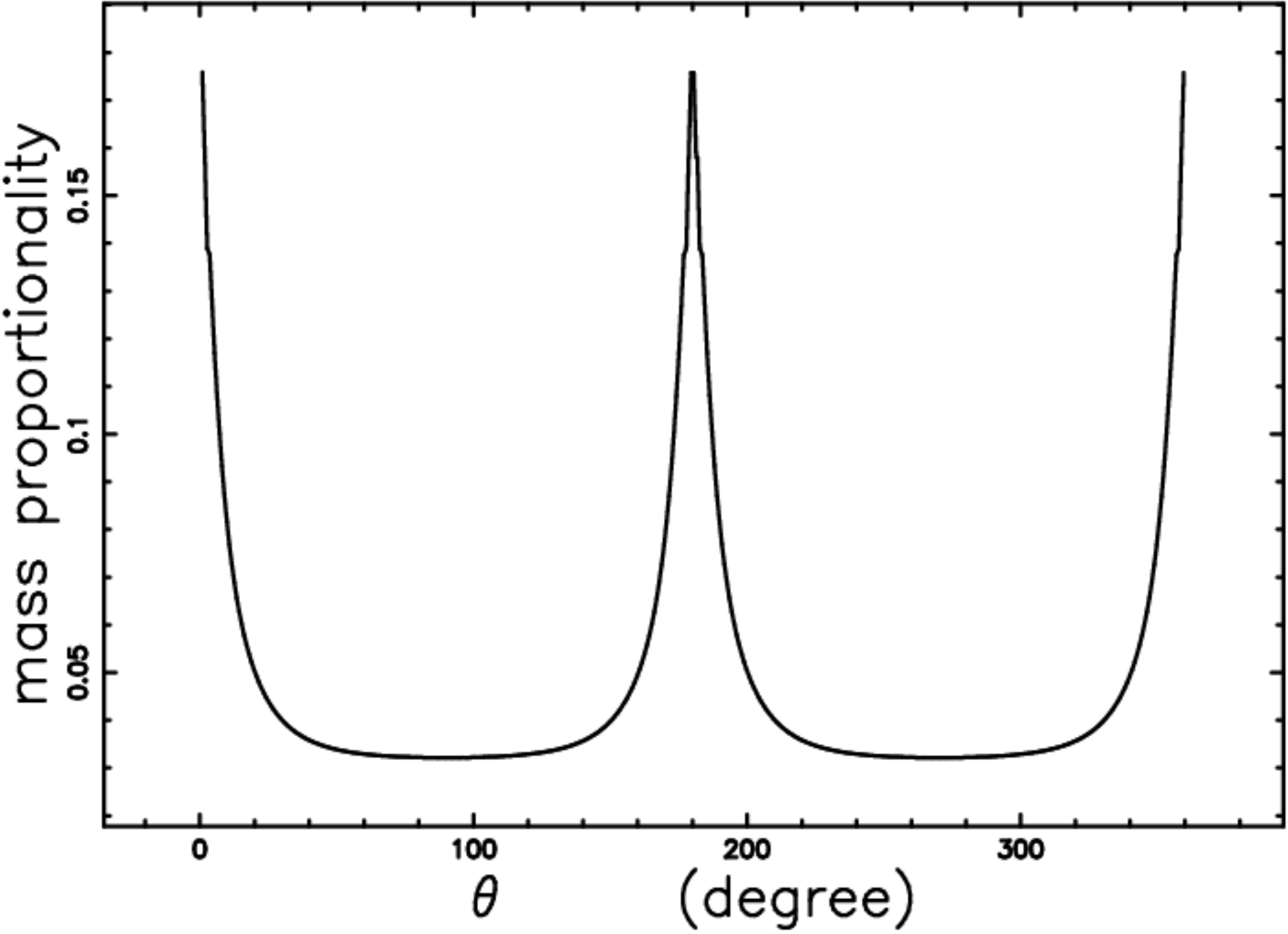}
\caption {
Swept mass   of \sn1987a as a
function of the position angle
in degrees in arbitrary units.
Physical parameters as
in Table \ref{datafit1987a}.
          }%
    \label{1987a_massa}
    \end{figure}

An explanation for the lack of velocity in the equatorial region
of  \sn1987a can be drawn from a careful analysis of Fig.
\ref{1987a_massa} which  displays the swept mass as a function of
the latitude. The ratio between maximum swept mass in the polar
direction and minimum swept mass in the equatorial plane is $
\approx $ 5.5  .  On applying the momentum conservation the
velocity in the equatorial plane is  5.5 times smaller in respect
to the polar direction where the velocity is maximum. This
numerical evaluation gives a  simple explanation for the asymmetry
of \snr : a smaller mass being swept means a greater velocity of
the advancing radius of the nebula.

\subsection{Results on the weakly asymmetric  \s1006}

The  input data   of the simulation are reported in
Table \ref{datafit1006}  ; the  reliability is
$\epsilon=89 \%$ for the radii  in the equatorial direction
and  $\epsilon=89 \%$ for the velocity in the equatorial
direction.
\begin{table}
\caption
{
Numerical value of the parameters
of the simulation  for  \s1006.
}
\label{datafit1006}
 \[
 \begin{array}{lc}
 \hline
 \hline
 R_{0,pc}         &   0.09  \\
\dot {R}_{0,kms}  &   8500  \\
p                 &   30    \\
t_{0,1}           &   10    \\
t_{1}             &   1000  \\
\noalign{\smallskip}
 \hline
 \hline
 \end{array}
 \]
 \end {table}

\begin{figure}
\includegraphics[width=6cm]{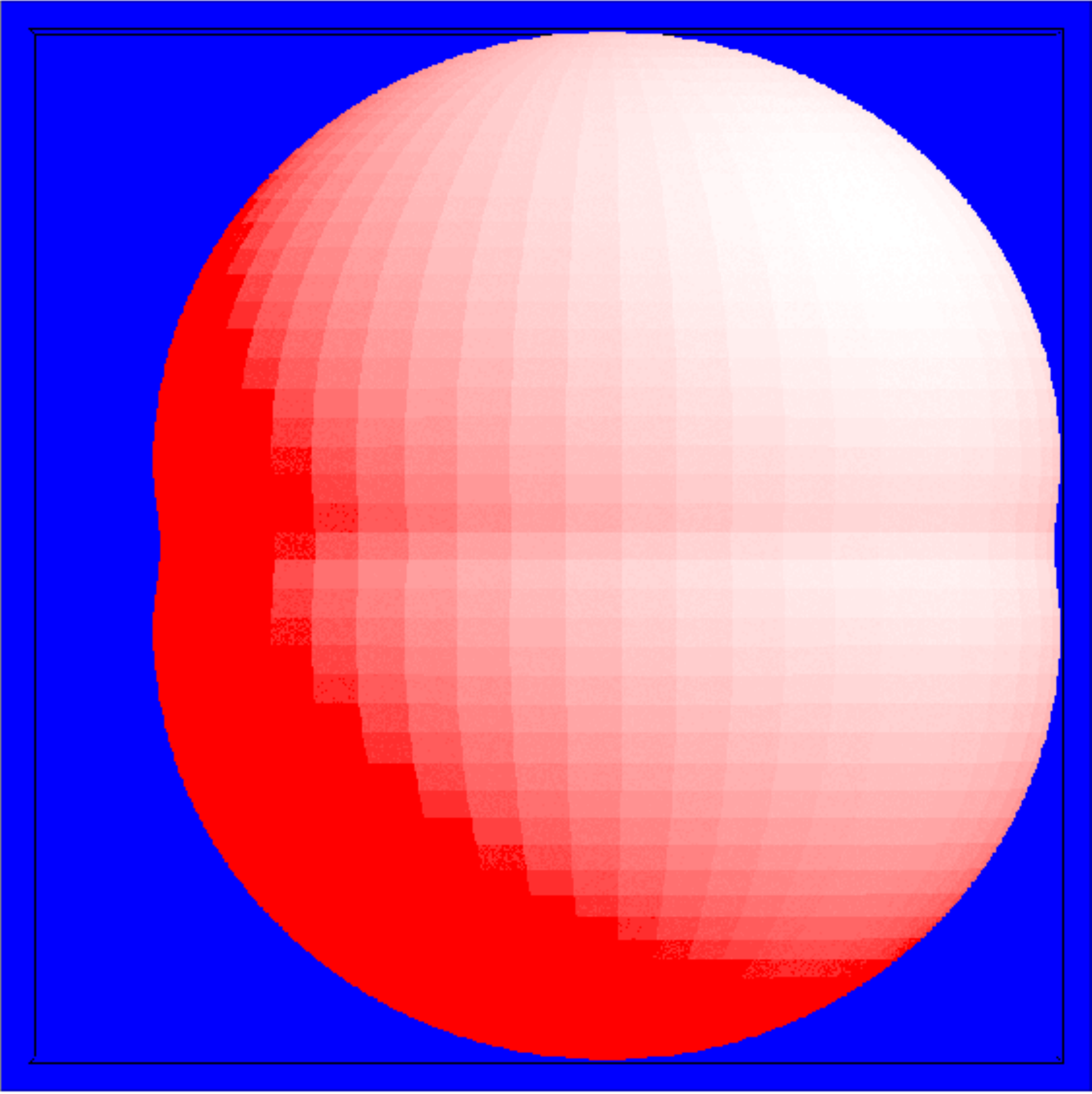}
\caption {
Continuous  three-dimensional surface of the
\s1006 : the three Eulerian angles
characterizing the point of
view are
     $ \Phi   $=90    $^{\circ }  $,
     $ \Theta $=90    $^{\circ }  $
and  $ \Psi   $=90    $^{\circ }  $.
Physical parameters as in Table \ref{datafit1987a}.
          }%
    \label{1006_faces}
    \end{figure}

The weakly asymmetric 3D shape  of  \s1006
is reported  in Fig. \ref{1006_faces}.

\begin{figure}
\includegraphics[width=6cm]{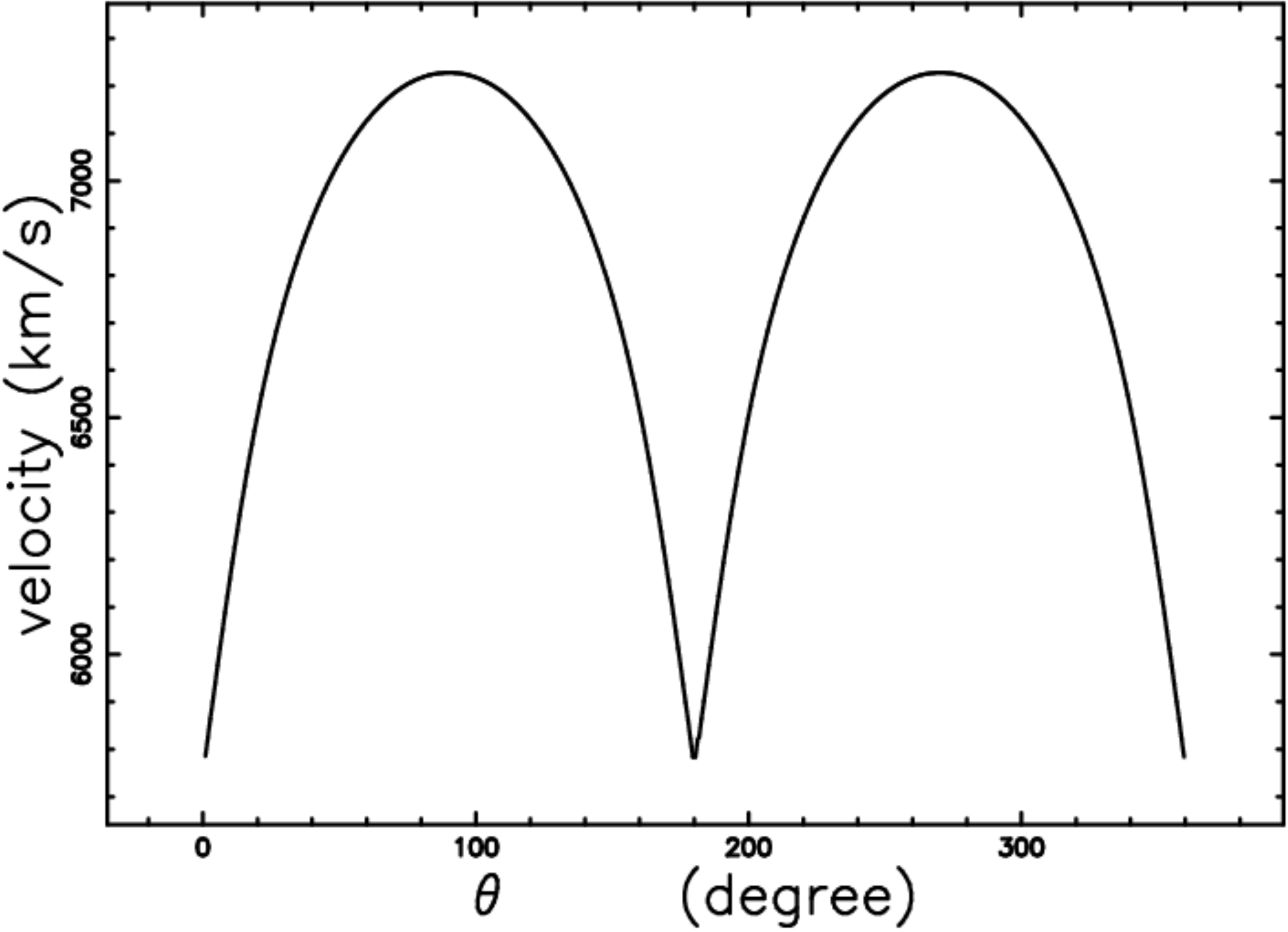}
\caption {
Velocity  in $\frac{km}{s}$  of \s1006 as a
function of the position angle in degrees.
Physical parameters as
in Table \ref{datafit1987a}.
          }%
    \label{1006_velocity}
    \end{figure}

The velocity  as  function  of the position angle
is  plotted  in Fig. \ref{1006_velocity}
and a comparison  should be done
with Fig. 4  in \cite{Katsuda2009}
where the proper motion as function of the azimuth angle
was reported.

\begin{figure}
\includegraphics[width=6cm]{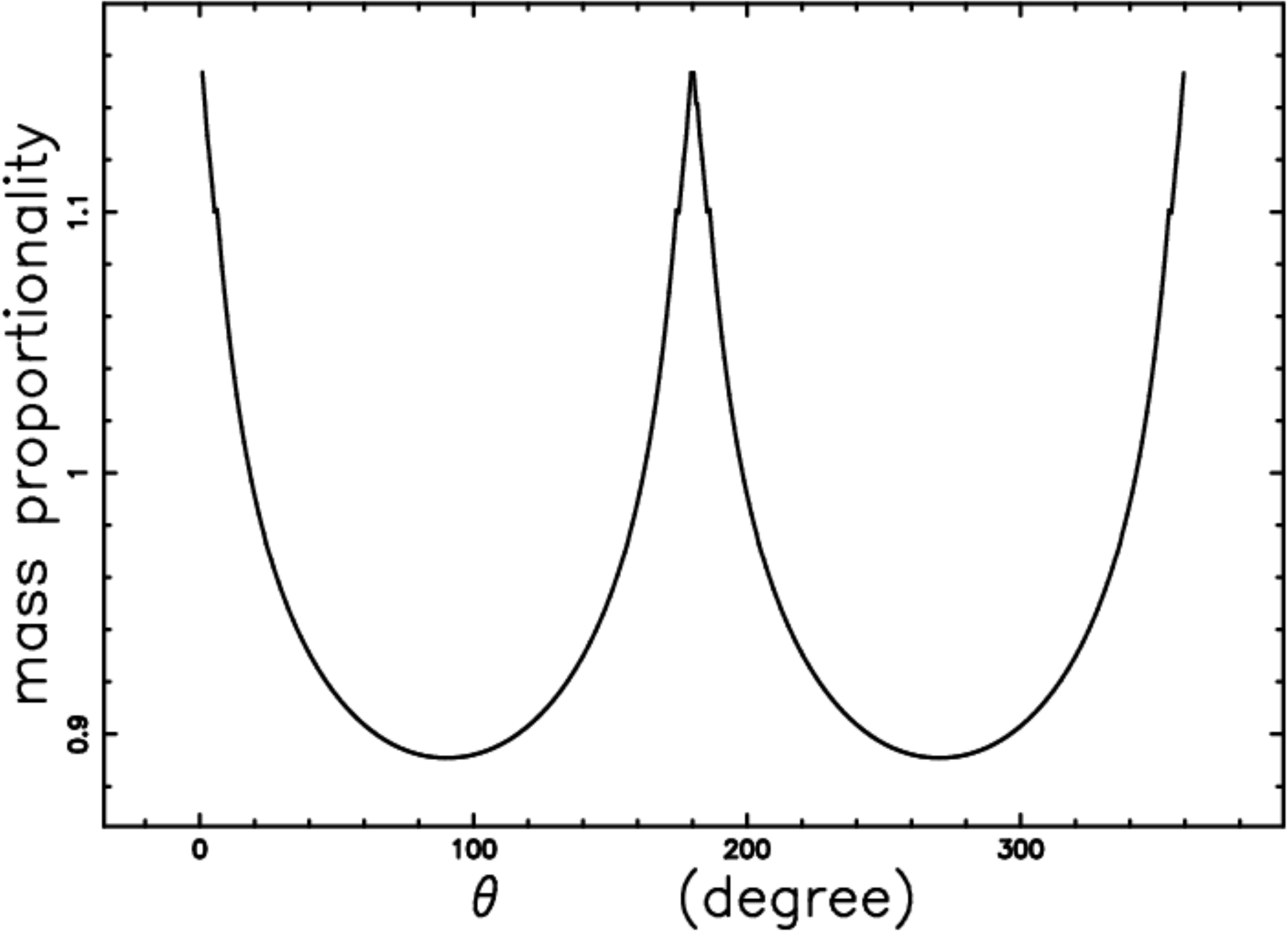}
\caption { Swept mass   of \s1006 as a function of the position
angle in degrees in arbitrary units. Physical parameters as in
Table \ref{datafit1006}.
          }%
    \label{1006_massa}
    \end{figure}

Our model for  \s1006 predicts a minimum velocity  in the
equatorial plane  of 5785 $km/s$  and  a maximum velocity of 7229
$km/s$  in the polar direction. A recent observation of \s1006
quotes a minimum velocity of
 5500 $km/s$  and a maximum velocity of 14500
$km/s$   assuming a distance of = 3.4 kpc
 The swept  mass
in the thin layer versus the  position angle is displayed in
  Fig.
\ref{1006_massa}.

\subsection{Comparison with the stellar wind}
\begin{figure}
\includegraphics[width=6cm]{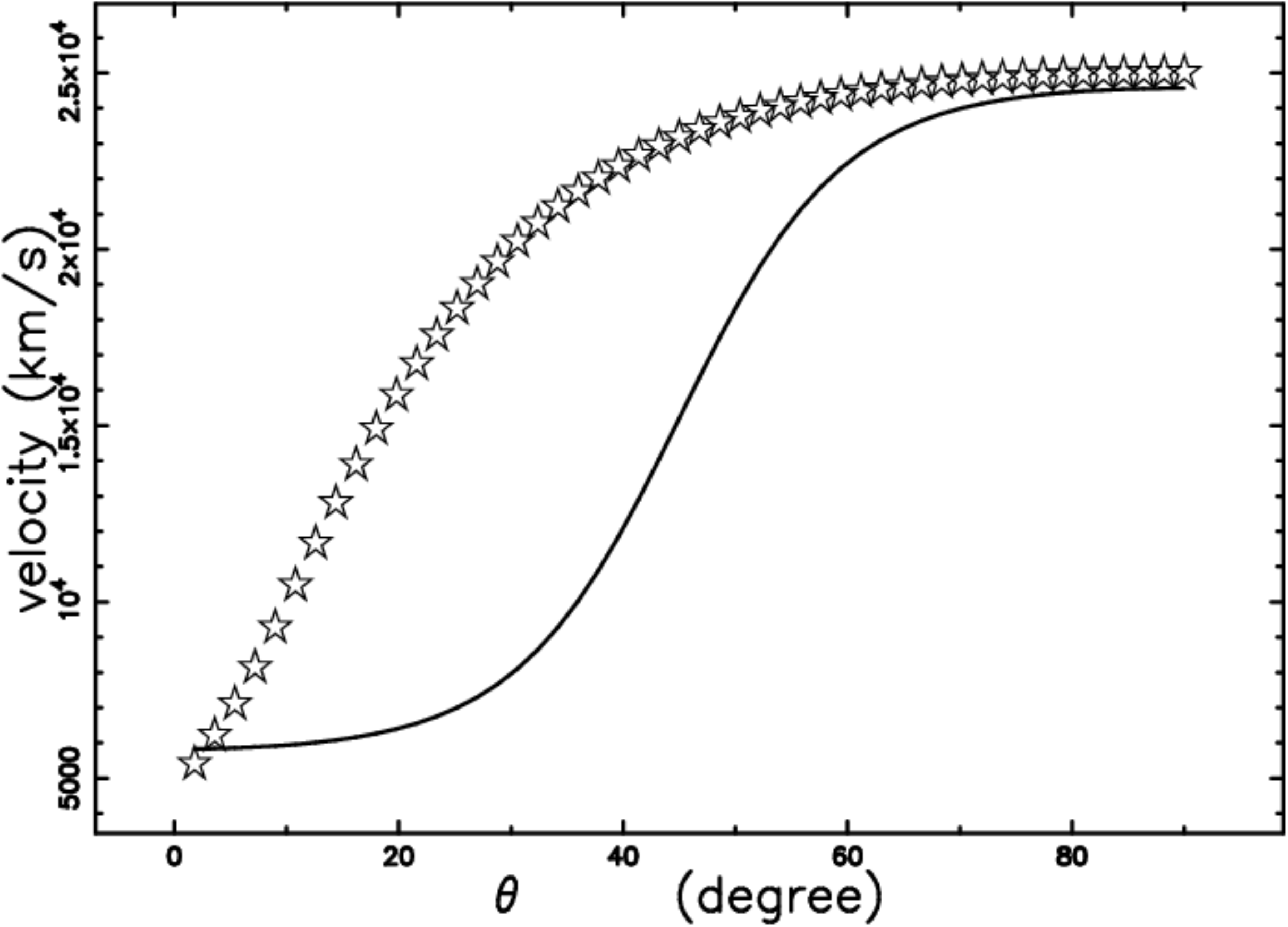}
\caption { Expansion velocity  versus latitude of  \sn1987a  (
open stars) and fit as given by  formula~(\ref{vgonzales}) (full
line). The fitting parameters are : $\lambda$=1.9,
$v_1=2503.6\,km/s$ and $v_2=5401\, km/s$.
          }%
    \label{1987a_velocity_gonzales}
    \end{figure}

A comparison can be done with the expansion  speed of the outer
$H_2$ shell of $\eta$-Carinae which has  been fitted with the
following latitude dependent velocity
\begin{equation}
v= \frac{ {\it v_1}\, \left( {\it v_2}+{{\rm e}^{2\,{\it
\lambda}\,\cos \left( 2\, \Theta \right) }}{\it v_1} \right) } {
{\it v_1}\, \left( 1+{{\rm e}^{2\,{\it \lambda}\,\cos \left(
2\,\Theta
 \right) }} \right)
} \quad, \label{vgonzales}
\end{equation}
where the parameter $\lambda$ controls the shape of the
Homunculus, $\Theta$ is the polar angle; $v_1$ and $v_2$ are the
velocities in the polar and equatorial direction, see
\cite{Gonzales2010}. Fig. \ref{1987a_velocity_gonzales} the data
of our simulation  as well the wind-type profile of velocity.

\section{Radiative transfer equation}

\label{sec_transfer}
The transfer equation in the presence of emission
only , see for
example \cite{rybicki} or \cite{Hjellming1988}, is
 \begin{equation}
\frac {dI_{\nu}}{ds} =  -k_{\nu} \zeta I_{\nu}  + j_{\nu} \zeta
\label{equazionetrasfer} \quad ,
\end {equation}
where  $I_{\nu}$ is the specific intensity , $s$  is the line of
sight , $j_{\nu}$ the emission coefficient, $k_{\nu}$   a mass
absorption coefficient, $\zeta$ the  mass density
at position $s$
and the index $\nu$ denotes the interested
frequency of emission.
The solution to  equation~(\ref{equazionetrasfer})
 is
\begin{equation}
 I_{\nu} (\tau_{\nu}) =
\frac {j_{\nu}}{k_{\nu}} ( 1 - e ^{-\tau_{\nu}(s)} ) \quad  ,
\label{eqn_transfer}
\end {equation}
where $\tau_{\nu}$ is the optical depth at frequency $\nu$
\begin{equation}
d \tau_{\nu} = k_{\nu} \zeta ds \quad.
\end {equation}
We now continue analyzing the case of an
 optically thin layer
in which $\tau_{\nu}$ is very small
( or $k_{\nu}$  very small )
and the density  $\zeta$ is substituted with our number density
$C(s)$ of  particles.
One case is taken into account :   the
emissivity is proportional to the number density.
\begin{equation}
j_{\nu} \zeta =K  C(s) \quad  ,
\end{equation}
where $K$ is a  constant function. This can be the case of
synchrotron radiation in presence of a  isotropic distribution of
electrons with a power law distribution in energy, $N(E)$,
\begin{equation}
N(E)dE = K_s E^{-\gamma_f} \label{spectrum} \quad  ,
\end{equation}
where $K_s$ is a constant. In this case  the emissivity is
\begin{eqnarray}
j_{\nu} \rho    \approx 0.933 \times 10^{-23} \alpha (\gamma_f)
K_s
 H_{\perp}
^{\frac{\gamma_f +1}{2} }  \times \nonumber \\
 \bigl (
 \frac{6.26 \times 10^{18} }{\nu}
\bigr )^{\frac{\gamma_f -1}{2} } \frac {erg} {s  cm^3  Hz
 rad^2} ,
\end{eqnarray}
where $\nu$ is the frequency and   $\alpha (\gamma_f)$  is a
slowly varying function of $\gamma_f$ which is of the order of
unity and is given by
\begin{eqnarray}
\alpha(\gamma_f) = \nonumber  \\
  2^{(\gamma_f -3)/2}
\frac{\gamma_f+7/3}{\gamma_f +1} \Gamma \bigl ( \frac {3\gamma_f
-1 }{12} \bigr ) \Gamma \bigl ( \frac {3\gamma_f +7 }{12} \bigr )
\quad ,
\end{eqnarray}
for  $\gamma_f \ge \frac{1}{2}$,
 see formula
(1.175 ) in  \cite{lang}.
The source of synchrotron luminosity
is assumed here to be
the flux of kinetic energy,
$L_m$,
\begin{equation}
L_m = \frac{1}{2}\rho A  V^3
\quad,
\label{fluxkineticenergy}
\end{equation}
where $A$ is the considered area, see formula (A28)
in \cite{deyoung}.
In our  case $A=4\pi R^2$,
which means
\begin{equation}
L_m = \frac{1}{2}\rho 4\pi R^2 V^3
\quad ,
\label{fluxkinetic}
\end{equation}
where $R$  is the instantaneous radius of the SNR and
$\rho$  is the density in the advancing layer
in which the synchrotron emission takes place.
The  total observed
luminosity
can  be expressed as
\begin{equation}
L_{tot} = \epsilon  L_{m}
\label{luminosity}
\quad  ,
\end{equation}
where  $\epsilon$  is  a constant  of conversion
from  the mechanical luminosity   to  the
total observed luminosity in synchrotron emission.
The fraction  of the total  luminosity
deposited  in a
given band   $f_b$  is
\begin{equation}
f_b  =
\frac
{
{{\it \nu_{b,min}}}^{-\gamma_f+1}-{{\it \nu_{b,max}}}^{-\gamma_f+1}
}
{
{{\it \nu_{min}}}^{-\gamma_f+1}-{{\it \nu_{max}}}^{-\gamma_f+1}
}
\quad  ,
\end{equation}
where  $\nu_{b,min}$  and  $\nu_{b,max}$
are the minimum and maximum frequency  of the given band.

\section{Image}

\label{sec_image}
An simulated image of an astrophysical object
is composed
by combining the intensities
which  characterize
different points.
For an optically thin medium the transfer equation
provides the emissivity to be multiplied with the distance
on  the line of sight , $l$.
This   length  in astrophysical diffuse objects
depends
on the  orientation of the observer.
A thermal and a non thermal model  are reviewed in the
spherical case.
In the the aspherical case a non thermal model is presented.

\subsection{Spherical  Image}
\begin{figure*}
\includegraphics[width=6cm]{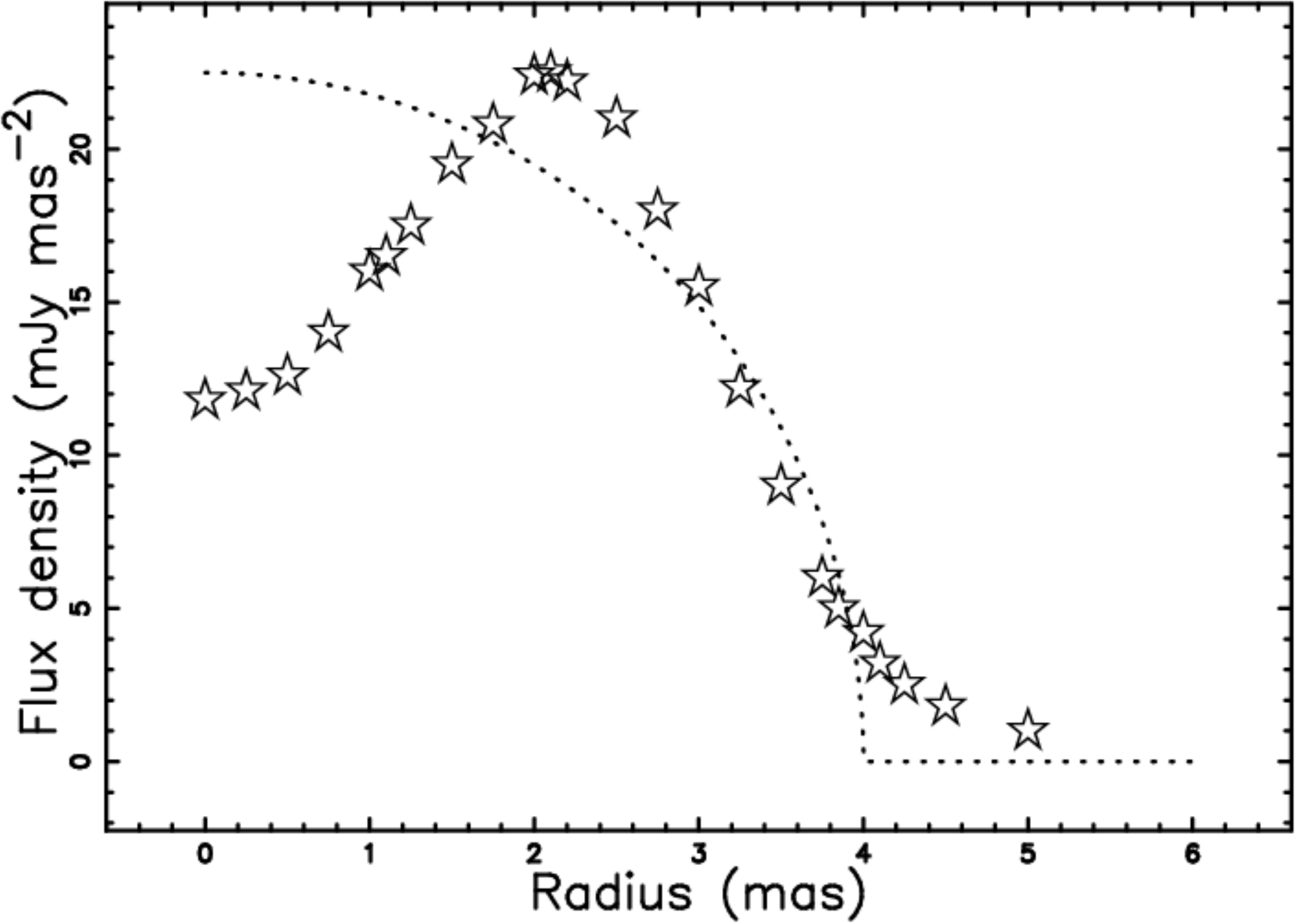}
\caption {
 Cut of the thermal   intensity ${\it I}$
 of the rim  model ( equation~(\ref{isphere}))
 through the center  (dotted  line) of \snr
 and  real data  (empty stars)
 when  $b=0.07$pc.
 The observed data as   day  1889 after  the explosion
 have been extracted  by the author
 from Fig.  3 of
 \cite{Marcaide2009}.
} \label{prof1993j_thermal}
    \end{figure*}

A {\it first} thermal model for the image   is characterized by a
constant temperature in  the internal region of the advancing
sphere. We therefore assume that the number density $C$ is
constant in a sphere of radius $a$ and then falls  to 0. The
length of sight , when the observer is situated at the infinity of
the $x$-axis , is the locus parallel to the $x$-axis which
crosses  the position $y$ in a Cartesian $x-y$ plane and
terminates at the external circle of radius $a$, see
\cite{Zaninetti2009a}. The locus  length is
\begin{eqnarray}
l_{ab} = 2 \times ( \sqrt {a^2 -y^2}) \quad  ;   0 \leq y < a
\quad . \label{lengthsphere}
\end{eqnarray}
The number density $C_m$ is constant  in the sphere of radius $a$
and therefore the intensity of radiation is
\begin{eqnarray}
I_{0a} =C_m \times  2 \times ( \sqrt { a^2 -y^2})
 \quad  ;  0 \leq y < a    \quad .
\label{isphere}
\end{eqnarray}
The comparison   of observed data of \snr  and the theoretical
thermal intensity is reported in Fig. \ref{prof1993j_thermal}.

\begin{figure*}
\includegraphics[width=6cm]{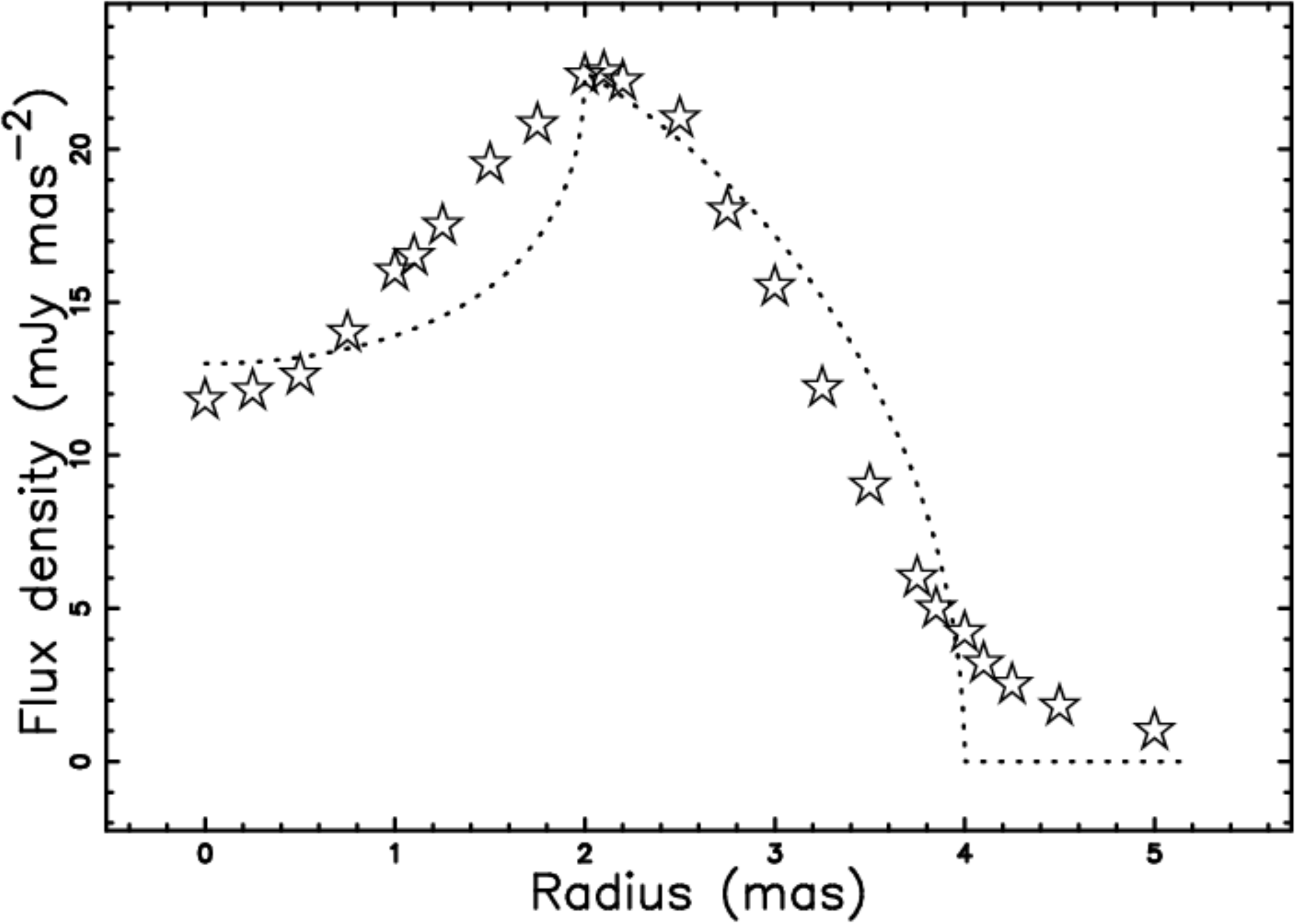}
\caption {
 Cut of the non thermal   intensity ${\it I}$
 of the rim  model ( equation~(\ref{irim}))
 through the center  (dotted  line) of \snr
 and  real data  (empty stars).
 The parameters  are
 $a=0.0035  $ pc and $b=0.07$pc.
Observed data as   day  1889 after  the explosion have been
extracted  by the author from Fig.  3 of Marcaide et al.   (2009).
} \label{prof1993j_const}
    \end{figure*}

A {\it second} non thermal model for the image is characterized by
emission in a thin layer around the advancing sphere. We therefore
assume that the number density $C$ is constant and in particular
rises from 0 at $r=a$ to a maximum value $C_m$ , remains constant
up to $r=b$ and then falls again to 0. The length of sight , when
the observer is situated at the infinity of the $x$-axis , is the
locus parallel to the $x$-axis which  crosses  the position $y$ in
a Cartesian $x-y$ plane and terminates at the external circle of
radius $b$, see \cite{Zaninetti2009a}.
 The locus length
is
\begin{eqnarray}
l_{0a} = 2 \times ( \sqrt { b^2 -y^2} - \sqrt {a^2 -y^2})
\quad  ;   0 \leq y < a  \nonumber  \\
l_{ab} = 2 \times ( \sqrt { b^2 -y^2})
 \quad  ;  a \leq y < b    \quad .
\label{length}
\end{eqnarray}
The number density $C_m$ is constant between two spheres of radius
$a$ and $b$ and therefore the intensity of radiation is
\begin{eqnarray}
I_{0a} =C_m \times 2 \times ( \sqrt { b^2 -y^2} - \sqrt {a^2
-y^2})
\quad  ;   0 \leq y < a  \nonumber  \\
I_{ab} =C_m \times  2 \times ( \sqrt { b^2 -y^2})
 \quad  ;  a \leq y < b    \quad .
\label{irim}
\end{eqnarray}
The comparison   of observed data of \snr  and the theoretical non
thermal intensity is displayed  in Fig. ~\ref{prof1993j_const}.

The main result of this Section is that the intensity of the
thermal model which has  the maximum of the intensity at the
center of SNR does not  match with the observed profiles. The
observed profiles in the intensity have the maximum value at the
rim as predicted by the non thermal model.

\subsection{Aspherical Image}

The numerical algorithm which allows us to
build  a complex  image is now
outlined.
\begin{itemize}
\item An empty (value=0)
memory grid  ${\mathcal {M}} (i,j,k)$ which  contains
$NDIM^3$ pixels is considered
\item We  first  generate an
internal 3D surface by rotating the ideal image
 $180^{\circ}$
around the polar direction and a second  external  surface at a
fixed distance $\Delta R$ from the first surface. As an example,
we fixed $\Delta R$ = $ 0.03 R_{max}$, where $R_{max}$ is the
maximum radius of expansion.
The points on
the memory grid which lie between the internal and external
surfaces are memorized on
${\mathcal {M}} (i,j,k)$ with a variable integer
number   according to formula
(\ref{fluxkinetic})  and   density $\rho$ proportional
to the swept    mass, see  Fig. \ref{1987a_massa}.
\item Each point of
${\mathcal {M}} (i,j,k)$  has spatial coordinates $x,y,z$ which  can be
represented by the following $1 \times 3$  matrix, $A$,
\begin{equation}
A=
 \left[ \begin {array}{c} x \\\noalign{\medskip}y\\\noalign{\medskip}{
\it z}\end {array} \right]
\quad  .
\end{equation}
The orientation  of the object is characterized by
 the
Euler angles $(\Phi, \Theta, \Psi)$
and  therefore  by a total
 $3 \times 3$  rotation matrix,
$E$, see \cite{Goldstein2002}.
The matrix point  is
represented by the following $1 \times 3$  matrix, $B$,
\begin{equation}
B = E \cdot A
\quad .
\end{equation}
\item
The intensity map is obtained by summing the points of the
rotated images
along a particular direction.
\item
The effect of the  insertion of a threshold intensity
, $I_{tr}$,
given by the observational techniques ,
is now analyzed.
The threshold intensity can be
parametrized  to  $I_{max}$,
the maximum  value  of intensity
characterizing the map.
\end{itemize}

\subsection{The image of the strongly asymmetric \sn1987a}

\begin{figure}
\includegraphics[width=6cm]{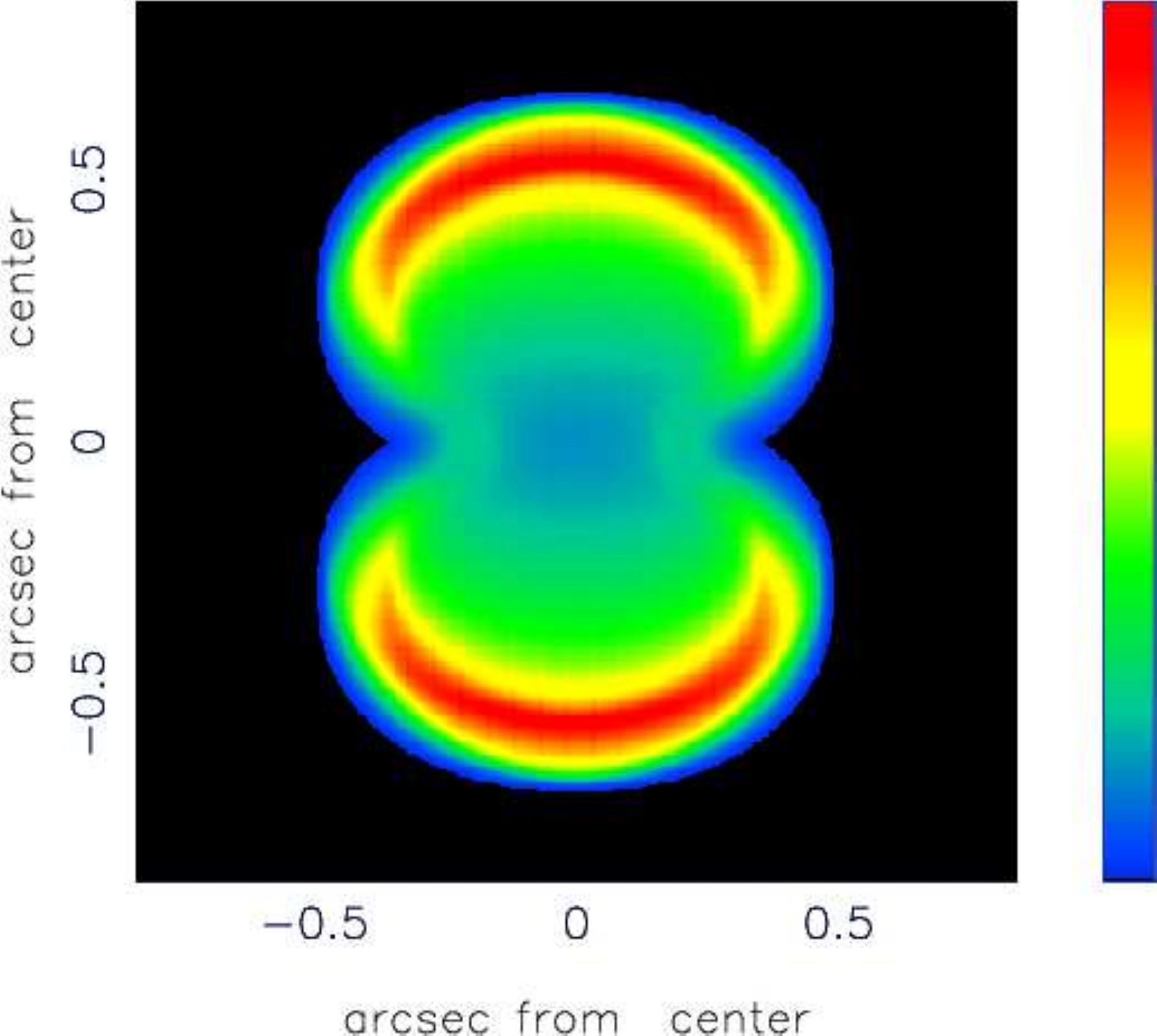}
\caption {
Map of the theoretical intensity  of
\sn1987a
in the presence of an exponentially varying medium.
Physical parameters as in Table \ref{datafit1987a}.
The three Euler angles
characterizing the   orientation
  are $ \Phi $=180$^{\circ }$,
$ \Theta     $=90 $^{\circ }$
and   $ \Psi $=0  $^{\circ }$.
This  combination of Euler angles corresponds
to the rotated image with the polar axis along the
z-axis.}%
    \label{1987a_heat}
    \end{figure}

\begin{figure}
\includegraphics[width=6cm]{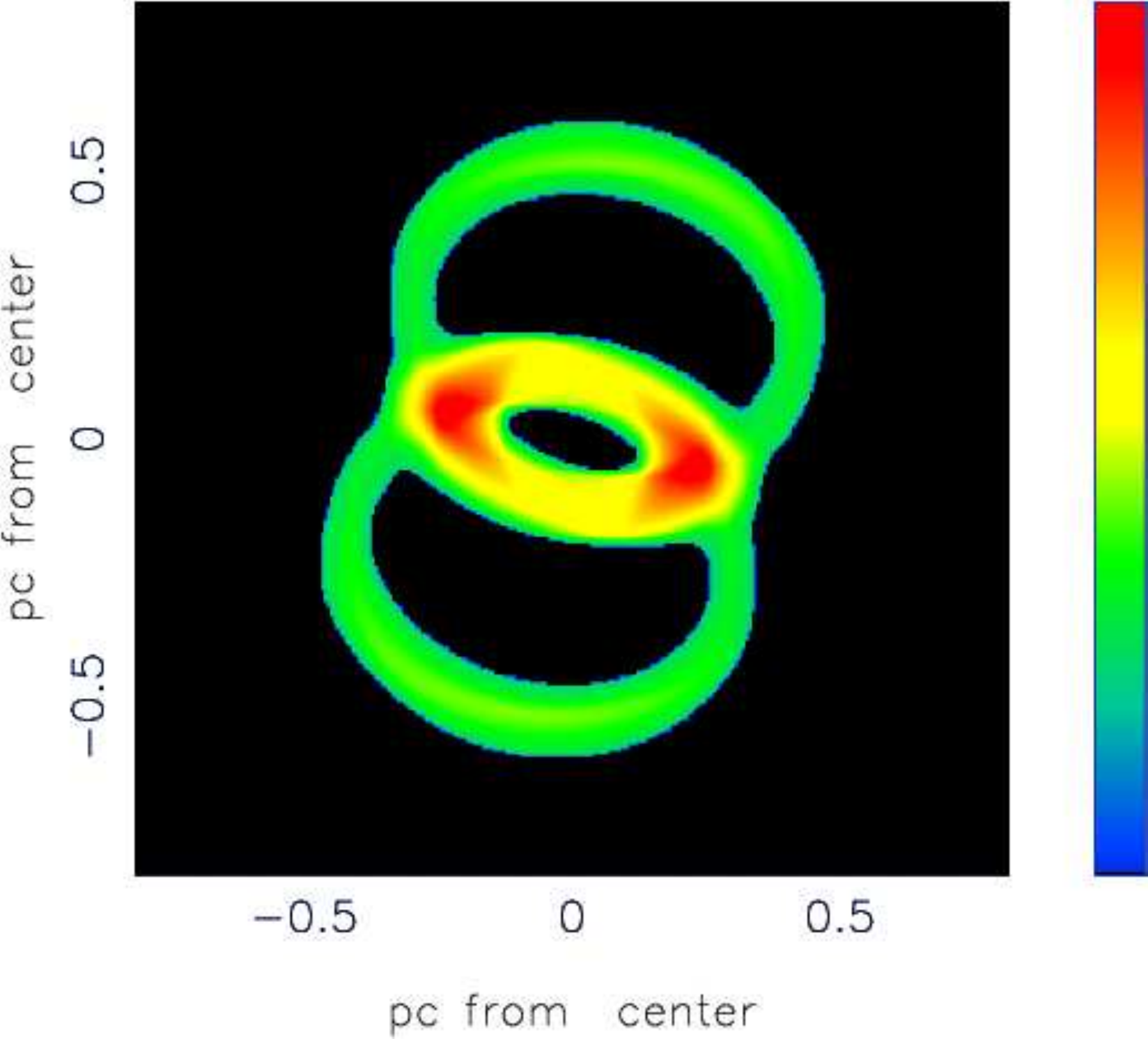}
\caption {
Model map of \sn1987a rotated in
accordance with the observations,
for an exponentially varying medium.
Physical parameters as in Table \ref{datafit1987a}.
The three Euler  angles characterizing
the orientation of the observer
are
     $ \Phi   $=105$^{\circ }$,
     $ \Theta $=55 $^{\circ }$
and  $ \Psi   $=-165 $^{\circ }$.
This  combination of Euler angles corresponds
to the observed image.
In this map $I_{tr}= I_{max}/4$
          }%
    \label{1987a_heat_hole_2}
    \end{figure}

An  ideal image of  \sn1987a
having the polar axis aligned with the z-direction
which means  polar axis along the z-direction,
is shown in Fig. \ref{1987a_heat}.
A model for a realistically rotated \sn1987a
is shown in Fig. \ref{1987a_heat_hole_2}.

The three rings of \sn1987a  are now simulated
in our Fig.  \ref{1987a_heat_hole_2}
and a comparison  should be done
with  Fig. 1 in \cite{Tziamtzis2011}
in which  the HST/ACS image (filter F658N) of the triple
ring system of \sn1987a was  reported.

\subsection{The image of the weakly asymmetric \s1006}

\begin{figure}
\includegraphics[width=6cm]{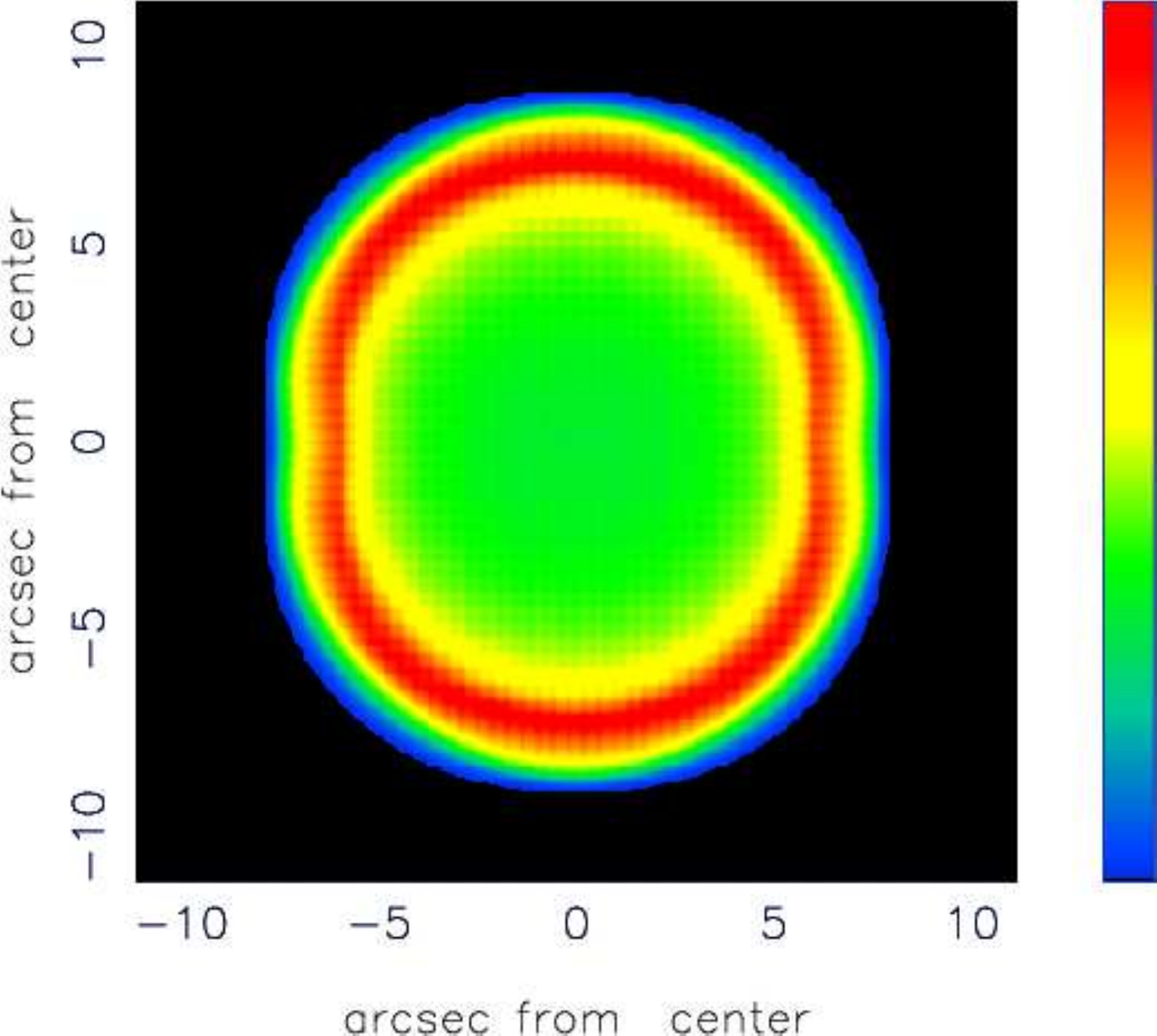}
\caption { Non rotated  map of \s1006   for an exponentially
varying medium. Physical parameters as in Table \ref{datafit1006}.
          }%
    \label{1006_heat}
    \end{figure}

\begin{figure*}
\includegraphics[width=6cm]{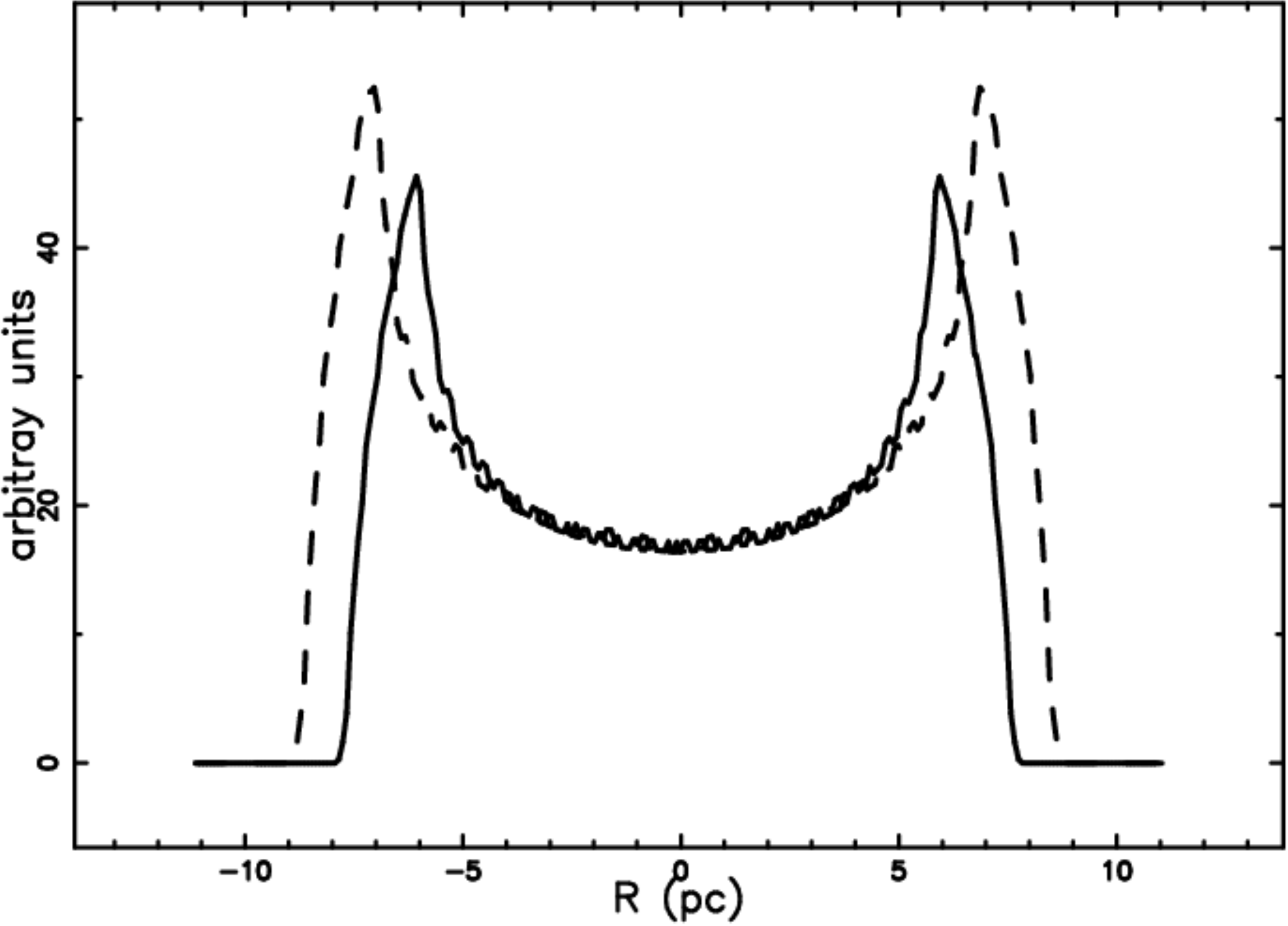}
\caption
{
 Two cut  along  perpendicular lines  of
 {\it I }   for the non rotated image  of SNR  \s1006.
}
\label{cut_xy_1006_lum}
    \end{figure*}

\begin{figure*}
\includegraphics[width=6cm]{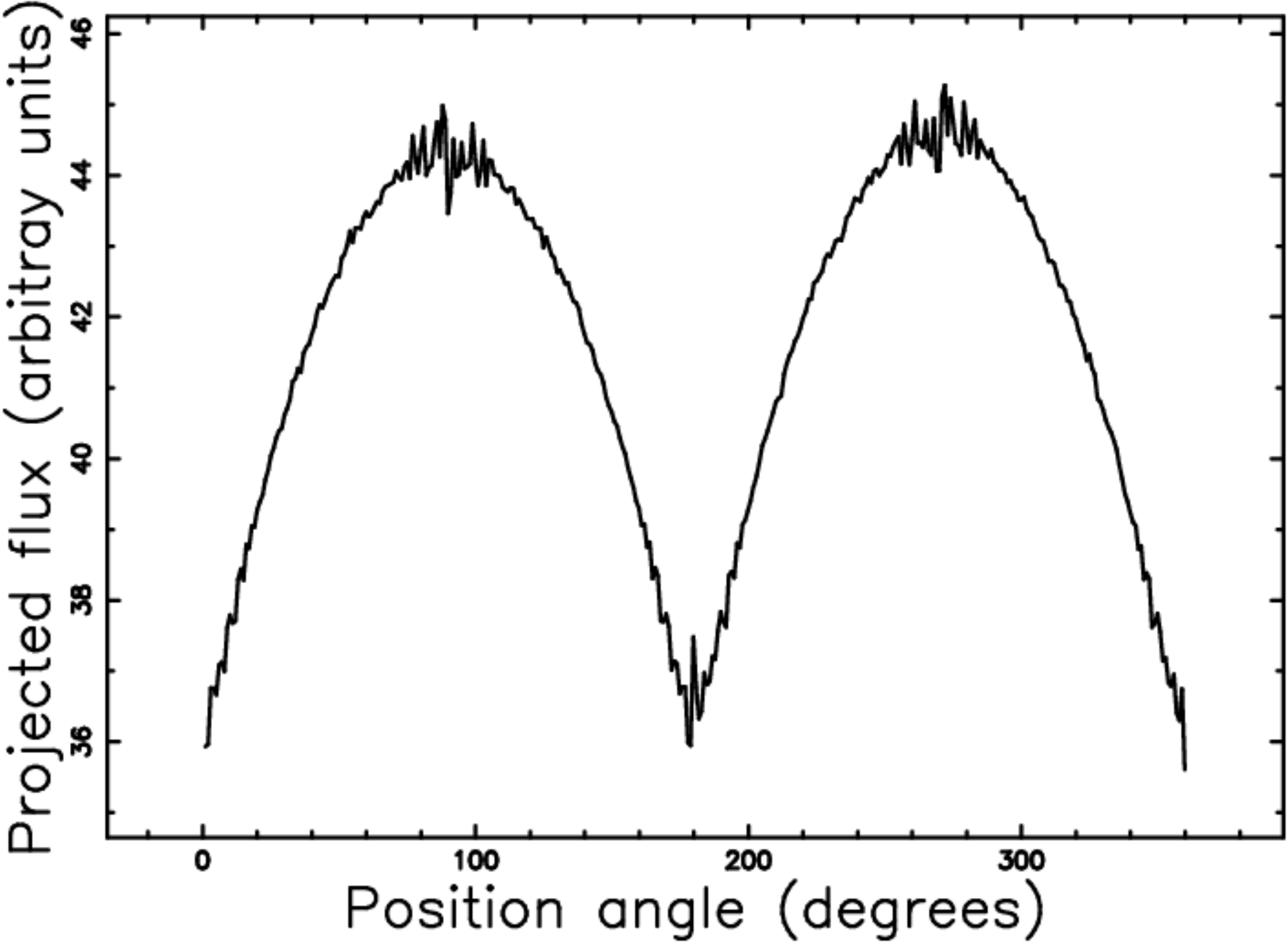}
\caption
{
Intensity
as function of the position angle
in degrees for SNR  \s1006.
}
\label{flux_sn1006_360}
    \end{figure*}

The image of  \s1006   is  visible in different bands  such as
radio, see \cite{Reynolds1993,Reynoso2007}, optical  , see
\cite{Long2007} and X-ray , see  \cite{Dyer,Katsuda2010}. The  2D
map in intensity of \s1006 is  visible in Fig. \ref{1006_heat}.
The  intensity along the equatorial and polar direction of our
image is reported in   Fig.  \ref{cut_xy_1006_lum}; a comparison
should be done with Fig.  4 in \cite{Dyer}.
The  projected flux  as a function of the  position angle
is another interesting quantity to  plot ,
see Fig.  \ref{flux_sn1006_360}
and a comparison should  be done
 with  Fig. 5 top right
in  \cite{Rothenflug2004}.

\begin{figure}
\includegraphics[width=6cm]{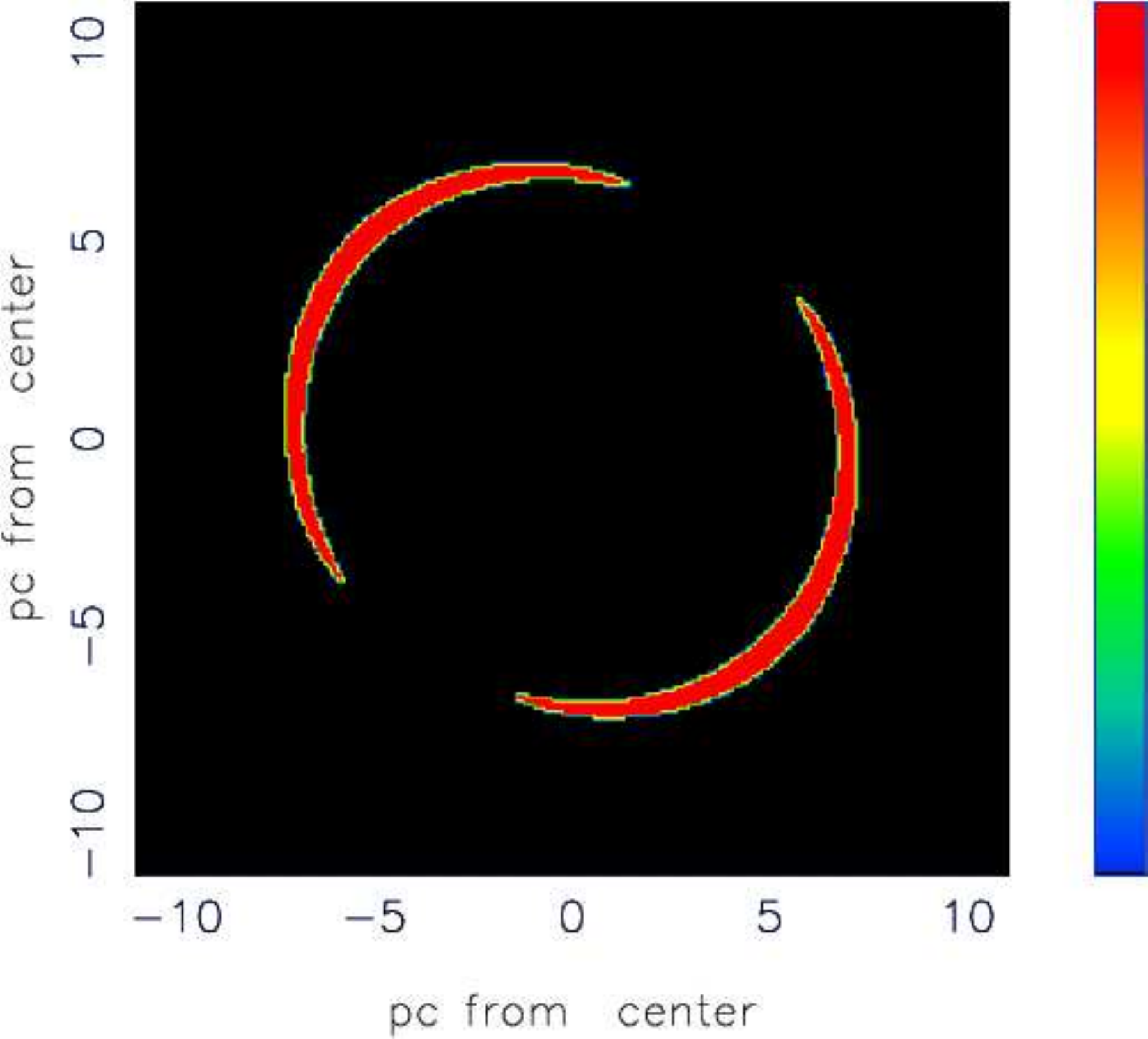}
\caption {
Model map of \s1006 rotated in
accordance with the X observations,
for an exponentially varying medium.
Physical parameters as in Table \ref{datafit1006}.
The three Euler  angles characterizing
the orientation of the observer
are
     $ \Phi   $=90   $^{\circ }$,
     $ \Theta $=-55  $^{\circ }$
and  $ \Psi   $=-180 $^{\circ }$.
This  combination of Euler angles corresponds
to the observed image.
In this map $I_{tr}= I_{max}/1.1$
          }%
    \label{1006_x}
    \end{figure}

After the previous graphs  is  more simple to  present a
characteristic feature  as  the "jet  appearance" visible in some
maps , see  our Fig. \ref{1006_x}; a comparison should be done
with the X-map  at 6.33-6.53  kev  band  visible  in Fig. 3b by
\cite{Yamaguchi2008}.

\section{Conclusions}

{\bf Law of motion}

We have deduced a new law  of motion
in spherical  symmetry ( constant density)
for an
advancing shell  assuming that only a fraction
of the mass  which  resides in the
surrounding medium  is  accumulated in the
advancing layer, see  equation (\ref{rtclassical}).
The presence  of an exponential  law for the density  transforms
the spherical symmetry  in axial  symmetry
and allows  the appearance  of
the so called "bipolar motion",
see  the  nonlinear  astrophysical equation
 (\ref{nonlinearastro}).

{\bf Images}

The emissivity in the advancing  layer is assumed
to be  proportional  to the  flux of  kinetic energy,
see  equation   (\ref{fluxkineticenergy})
where  the density  is assumed to be
proportional  to the swept material.
This  assumption  allows to simulate
particular effects such as
the triple ring system of \sn1987a, see
Fig. \ref{1987a_heat_hole_2}.
Another  curious  effect is  the  "jet  appearance" visible in the
weakly  symmetric \s1006, see  Fig. \ref{1006_x}. The jet/counter
jet effect plays a relevant role in the actual research , see
discussion in Section 5.2 in \cite{Dopita2006} where the  jet
appearance is tentatively explained by the neutrino heating , see
\cite{Walder2005} or  by the MHD jet , see \cite{Takiwaki2004}.
Here conversely we explain the  appearance  of the jet by the
addition of three effects :
 \begin{itemize}
\item An asymmetric law of expansion due to a gradient in density
in respect to the equatorial plane which produces an asymmetry in
velocity . \item The direct conversion of the flux of kinetic
energy into radiation.
 \item The image  of the SNR as the composition of integrals
 along the line of sight.
 \end{itemize}
 According to the previous three  ingredients the  neutrino
 heating mechanism  is not necessary. The Magneto Hydrodynamic
 (MHD) approach is   supposed to act  in a hidden way
 on scales smaller than
 than the thickness of the advancing layer in order to accelerate
 the electrons to relativistic  energies.
A careful  calibration  of the various  involved  parameters
can be done  when  cuts  in intensity  are available.


\begin{thebibliography}{40}
\ifx \bisbn   \undefined \def \bisbn  #1{ISBN #1}\fi
\ifx \binits  \undefined \def \binits#1{#1} \fi
\ifx \bauthor  \undefined \def \bauthor#1{#1} \fi
\ifx \batitle  \undefined \def \batitle#1{#1} \fi
\ifx \bjtitle  \undefined \def \bjtitle#1{#1}\fi
\ifx \bvolume  \undefined \def \bvolume#1{\textbf{#1}}\fi
\ifx \byear  \undefined \def \byear#1{#1} \fi
\ifx \bissue  \undefined \def \bissue#1{#1} \fi
\ifx \bfpage  \undefined \def \bfpage#1{#1} \fi
\ifx \blpage  \undefined \def \blpage #1{#1} \fi
\ifx \burl  \undefined \def \burl#1{\textsf{#1}} \fi
\ifx \doiurl  \undefined \def \doiurl#1{\textsf{#1}} \fi
\ifx \betal  \undefined \def \betal{\textit{et al.}} \fi
\ifx \binstitute  \undefined \def \binstitute#1{#1} \fi
\ifx \bctitle  \undefined \def \bctitle#1{#1} \fi
\ifx \beditor  \undefined \def \beditor#1{#1} \fi
\ifx \bpublisher  \undefined \def \bpublisher#1{#1} \fi
\ifx \bbtitle  \undefined \def \bbtitle#1{#1} \fi
\ifx \bedition  \undefined \def \bedition#1{#1} \fi
\ifx \bseriesno  \undefined \def \bseriesno#1{#1} \fi
\ifx \blocation  \undefined \def \blocation#1{#1} \fi
\ifx \bsertitle  \undefined \def \bsertitle#1{#1} \fi
\ifx \bsnm \undefined \def \bsnm#1{#1} \fi
\ifx \bsuffix \undefined \def \bsuffix#1{#1} \fi
\ifx \bparticle \undefined \def \bparticle#1{#1} \fi
\ifx \barticle \undefined \def \barticle#1{#1} \fi
\ifx \botherref \undefined \def \botherref #1{#1} \fi
\ifx \url \undefined \def \url#1{\textsf{#1}} \fi
\ifx \bchapter \undefined \def \bchapter#1{#1} \fi
\ifx \bbook \undefined \def \bbook#1{#1} \fi
\ifx \bcomment \undefined \def \bcomment#1{#1} \fi
\ifx \oauthor \undefined \def \oauthor#1{#1} \fi
\ifx \citeauthoryear \undefined \def \citeauthoryear#1{#1} \fi
\def \endbibitem {}

\bibitem[\protect\citeauthoryear{{Bamba} \textit{et~al.}}{2003}]{Bamba2003}
\begin{barticle}
\bauthor{\bsnm{{Bamba}}, \binits{A.}}, \bauthor{\bsnm{{Yamazaki}},
  \binits{R.}}, \bauthor{\bsnm{{Ueno}}, \binits{M.}}, \bauthor{\bsnm{{Koyama}},
  \binits{K.}}:
\bjtitle{\apj}
\bvolume{589},
\bfpage{827}
(\byear{2003})
\end{barticle}
\endbibitem

\bibitem[\protect\citeauthoryear{{Cant{\'o}}, {Raga}, and
  {Adame}}{2006}]{Canto2006}
\begin{barticle}
\bauthor{\bsnm{{Cant{\'o}}}, \binits{J.}}, \bauthor{\bsnm{{Raga}},
  \binits{A.C.}}, \bauthor{\bsnm{{Adame}}, \binits{L.}}:
\bjtitle{\mnras}
\bvolume{369},
\bfpage{860}
(\byear{2006})
\end{barticle}
\endbibitem

\bibitem[\protect\citeauthoryear{{de Young}}{2002}]{deyoung}
\begin{bbook}
\bauthor{\bsnm{{de Young}}, \binits{D.S.}}:
\bbtitle{{The physics of extragalactic radio sources}}.
\bpublisher{University of Chicago Press},
\blocation{Chicago}
(\byear{2002})
\end{bbook}
\endbibitem

\bibitem[\protect\citeauthoryear{{Dyer}, {Reynolds}, and
  {Borkowski}}{2004}]{Dyer}
\begin{barticle}
\bauthor{\bsnm{{Dyer}}, \binits{K.K.}}, \bauthor{\bsnm{{Reynolds}},
  \binits{S.P.}}, \bauthor{\bsnm{{Borkowski}}, \binits{K.J.}}:
\bjtitle{\apj}
\bvolume{600},
\bfpage{752}
(\byear{2004})
\end{barticle}
\endbibitem

\bibitem[\protect\citeauthoryear{{Dyson}}{1983}]{Dyson1983}
\begin{barticle}
\bauthor{\bsnm{{Dyson}}, \binits{J.E.}}:
\bjtitle{\aap}
\bvolume{124},
\bfpage{77}
(\byear{1983})
\end{barticle}
\endbibitem

\bibitem[\protect\citeauthoryear{{{Dyson}, J.~E. and {Williams},
  D.~A.}}{1997}]{Dyson1997}
\begin{bbook}
\bauthor{\bsnm{{{Dyson}, J.~E. and {Williams}, D.~A.}}}:
\bbtitle{{The physics of the interstellar medium}}.
\bpublisher{Institute of Physics Publishing},
\blocation{Bristol}
(\byear{1997})
\end{bbook}
\endbibitem

\bibitem[\protect\citeauthoryear{{Ellison} \textit{et~al.}}{1994}]{Ellison1994}
\begin{barticle}
\bauthor{\bsnm{{Ellison}}, \binits{D.C.}}, \bauthor{\bsnm{{Reynolds}},
  \binits{S.P.}}, \bauthor{\bsnm{{Borkowski}}, \binits{K.}},
  \bauthor{\bsnm{{Chevalier}}, \binits{R.}}, \bauthor{\bsnm{{Cox}},
  \binits{D.P.}}, \bauthor{\bsnm{{Dickel}}, \binits{J.R.}},
  \bauthor{\bsnm{{Pisarski}}, \binits{R.}}, \bauthor{\bsnm{{Raymond}},
  \binits{J.}}, \bauthor{\bsnm{{Spangler}}, \binits{S.R.}},
  \bauthor{\bsnm{{Volk}}, \binits{H.J.}}, \bauthor{\bsnm{{Wefel}},
  \binits{J.P.}}:
\bjtitle{\pasp}
\bvolume{106},
\bfpage{780}
(\byear{1994})
\end{barticle}
\endbibitem

\bibitem[\protect\citeauthoryear{{Eriksen} \textit{et~al.}}{2009}]{Eriksen2009}
\begin{barticle}
\bauthor{\bsnm{{Eriksen}}, \binits{K.A.}}, \bauthor{\bsnm{{Arnett}},
  \binits{D.}}, \bauthor{\bsnm{{McCarthy}}, \binits{D.W.}},
  \bauthor{\bsnm{{Young}}, \binits{P.}}:
\bjtitle{\apj}
\bvolume{697},
\bfpage{29}
(\byear{2009})
\end{barticle}
\endbibitem

\bibitem[\protect\citeauthoryear{{Fesen} \textit{et~al.}}{2006}]{Dopita2006}
\begin{barticle}
\bauthor{\bsnm{{Fesen}}, \binits{R.A.}}, \bauthor{\bsnm{{Hammell}},
  \binits{M.C.}}, \bauthor{\bsnm{{Morse}}, \binits{J.}},
  \bauthor{\bsnm{{Chevalier}}, \binits{R.A.}}, \bauthor{\bsnm{{Borkowski}},
  \binits{K.J.}}, \bauthor{\bsnm{{Dopita}}, \binits{M.A.}},
  \bauthor{\bsnm{{Gerardy}}, \binits{C.L.}}, \bauthor{\bsnm{{Lawrence}},
  \binits{S.S.}}, \bauthor{\bsnm{{Raymond}}, \binits{J.C.}},
  \bauthor{\bsnm{{van den Bergh}}, \binits{S.}}:
\bjtitle{\apj}
\bvolume{645},
\bfpage{283}
(\byear{2006}).
doi:\doiurl{10.1086/504254}
\end{barticle}
\endbibitem

\bibitem[\protect\citeauthoryear{{Goldstein}, {Poole}, and
  {Safko}}{2002}]{Goldstein2002}
\begin{bbook}
\bauthor{\bsnm{{Goldstein}}, \binits{H.}}, \bauthor{\bsnm{{Poole}},
  \binits{C.}}, \bauthor{\bsnm{{Safko}}, \binits{J.}}:
\bbtitle{{Classical mechanics}}.
\bpublisher{Addison-Wesley},
\blocation{San Francisco}
(\byear{2002})
\end{bbook}
\endbibitem

\bibitem[\protect\citeauthoryear{{Gonz{\'a}lez}
  \textit{et~al.}}{2010}]{Gonzales2010}
\begin{barticle}
\bauthor{\bsnm{{Gonz{\'a}lez}}, \binits{R.F.}}, \bauthor{\bsnm{{Villa}},
  \binits{A.M.}}, \bauthor{\bsnm{{G{\'o}mez}}, \binits{G.C.}},
  \bauthor{\bsnm{{de Gouveia Dal Pino}}, \binits{E.M.}},
  \bauthor{\bsnm{{Raga}}, \binits{A.C.}}, \bauthor{\bsnm{{Cant{\'o}}},
  \binits{J.}}, \bauthor{\bsnm{{Vel{\'a}zquez}}, \binits{P.F.}},
  \bauthor{\bsnm{{de La Fuente}}, \binits{E.}}:
\bjtitle{\mnras}
\bvolume{402},
\bfpage{1141}
(\byear{2010})
\end{barticle}
\endbibitem

\bibitem[\protect\citeauthoryear{{{Hjellming}, R.~M.}}{1988}]{Hjellming1988}
\begin{bbook}
\bauthor{\bsnm{{{Hjellming}, R.~M.}}}:
\bbtitle{{Radio stars IN Galactic and Extragalactic Radio Astronomy }}.
\bpublisher{Springer},
\blocation{New York}
(\byear{1988})
\end{bbook}
\endbibitem

\bibitem[\protect\citeauthoryear{{Katsuda} \textit{et~al.}}{2009}]{Katsuda2009}
\begin{barticle}
\bauthor{\bsnm{{Katsuda}}, \binits{S.}}, \bauthor{\bsnm{{Petre}}, \binits{R.}},
  \bauthor{\bsnm{{Long}}, \binits{K.S.}}, \bauthor{\bsnm{{Reynolds}},
  \binits{S.P.}}, \bauthor{\bsnm{{Winkler}}, \binits{P.F.}},
  \bauthor{\bsnm{{Mori}}, \binits{K.}}, \bauthor{\bsnm{{Tsunemi}},
  \binits{H.}}:
\bjtitle{\apjl}
\bvolume{692},
\bfpage{105}
(\byear{2009})
\end{barticle}
\endbibitem

\bibitem[\protect\citeauthoryear{{Katsuda} \textit{et~al.}}{2010}]{Katsuda2010}
\begin{barticle}
\bauthor{\bsnm{{Katsuda}}, \binits{S.}}, \bauthor{\bsnm{{Petre}}, \binits{R.}},
  \bauthor{\bsnm{{Mori}}, \binits{K.}}, \bauthor{\bsnm{{Reynolds}},
  \binits{S.P.}}, \bauthor{\bsnm{{Long}}, \binits{K.S.}},
  \bauthor{\bsnm{{Winkler}}, \binits{P.F.}}, \bauthor{\bsnm{{Tsunemi}},
  \binits{H.}}:
\bjtitle{\apj}
\bvolume{723},
\bfpage{383}
(\byear{2010})
\end{barticle}
\endbibitem

\bibitem[\protect\citeauthoryear{{Lang}}{1999}]{lang}
\begin{bbook}
\bauthor{\bsnm{{Lang}}, \binits{K.R.}}:
\bbtitle{{Astrophysical formulae. (Third Edition)}}.
\bpublisher{Springer},
\blocation{New York}
(\byear{1999})
\end{bbook}
\endbibitem

\bibitem[\protect\citeauthoryear{{Long}}{2007}]{Long2007}
\begin{barticle}
\bauthor{\bsnm{{Long}}, \binits{K.S.}}:
\bjtitle{Highlights of Astronomy}
\bvolume{14},
\bfpage{306}
(\byear{2007})
\end{barticle}
\endbibitem

\bibitem[\protect\citeauthoryear{{Marcaide}
  \textit{et~al.}}{2009}]{Marcaide2009}
\begin{barticle}
\bauthor{\bsnm{{Marcaide}}, \binits{J.M.}},
  \bauthor{\bsnm{{Mart{\'{\i}}-Vidal}}, \binits{I.}},
  \bauthor{\bsnm{{Alberdi}}, \binits{A.}}, \bauthor{\bsnm{{P{\'e}rez-Torres}},
  \binits{M.A.}}:
\bjtitle{\aap}
\bvolume{505},
\bfpage{927}
(\byear{2009})
\end{barticle}
\endbibitem

\bibitem[\protect\citeauthoryear{{McCray} and {Layzer}}{1987}]{Dalgarno1987}
\begin{bbook}
\beditor{\bsnm{{McCray}}, \binits{A.} \bsuffix{R.~In:~{Dalgarno}}},
  \beditor{\bsnm{{Layzer}}, \binits{D.}} (eds.):
\bbtitle{{Spectroscopy of astrophysical plasmas}}.
\bpublisher{{Cambridge University Press}},
\blocation{Cambridge}
(\byear{1987})
\end{bbook}
\endbibitem

\bibitem[\protect\citeauthoryear{{McKee}}{1987}]{mckee}
\begin{botherref}
\oauthor{\bsnm{{McKee}}, \binits{C.F.}}:
In: {Dalgarno}, A., {Layzer}, D. (eds.)
Spectroscopy of Astrophysical Plasmas,
p. 226
(1987)
\end{botherref}
\endbibitem

\bibitem[\protect\citeauthoryear{{Mitchell}
  \textit{et~al.}}{2002}]{Mitchell2002}
\begin{barticle}
\bauthor{\bsnm{{Mitchell}}, \binits{R.C.}}, \bauthor{\bsnm{{Baron}},
  \binits{E.}}, \bauthor{\bsnm{{Branch}}, \binits{D.}},
  \bauthor{\bsnm{{Hauschildt}}, \binits{P.H.}}, \bauthor{\bsnm{{Nugent}},
  \binits{P.E.}}, \bauthor{\bsnm{{Lundqvist}}, \binits{P.}},
  \bauthor{\bsnm{{Blinnikov}}, \binits{S.}}, \bauthor{\bsnm{{Pun}},
  \binits{C.S.J.}}:
\bjtitle{\apj}
\bvolume{574},
\bfpage{293}
(\byear{2002})
\end{barticle}
\endbibitem

\bibitem[\protect\citeauthoryear{{Panagia}}{2005}]{Panagia2005}
\begin{botherref}
\oauthor{\bsnm{{Panagia}}, \binits{N.}}:
In: {Marcaide}, J.M., {Weiler}, K.W. (eds.)
IAU Colloq. 192: Cosmic Explosions, On the 10th Anniversary of SN1993J,
p. 585
(2005)
\end{botherref}
\endbibitem

\bibitem[\protect\citeauthoryear{{Petruk} and {Beshlei}}{2007}]{Petruk2007}
\begin{barticle}
\bauthor{\bsnm{{Petruk}}, \binits{O.}}, \bauthor{\bsnm{{Beshlei}},
  \binits{V.}}:
\bjtitle{Kinematics and Physics of Celestial Bodies}
\bvolume{23},
\bfpage{16}
(\byear{2007})
\end{barticle}
\endbibitem

\bibitem[\protect\citeauthoryear{{Press} \textit{et~al.}}{1992}]{press}
\begin{bbook}
\bauthor{\bsnm{{Press}}, \binits{W.H.}}, \bauthor{\bsnm{{Teukolsky}},
  \binits{S.A.}}, \bauthor{\bsnm{{Vetterling}}, \binits{W.T.}},
  \bauthor{\bsnm{{Flannery}}, \binits{B.P.}}:
\bbtitle{{Numerical Recipes in FORTRAN. The Art of Scientific Computing}}.
\bpublisher{Cambridge University Press},
\blocation{Cambridge}
(\byear{1992})
\end{bbook}
\endbibitem

\bibitem[\protect\citeauthoryear{{Racusin} \textit{et~al.}}{2009}]{Racusin2009}
\begin{barticle}
\bauthor{\bsnm{{Racusin}}, \binits{J.L.}}, \bauthor{\bsnm{{Park}},
  \binits{S.}}, \bauthor{\bsnm{{Zhekov}}, \binits{S.}},
  \bauthor{\bsnm{{Burrows}}, \binits{D.N.}}, \bauthor{\bsnm{{Garmire}},
  \binits{G.P.}}, \bauthor{\bsnm{{McCray}}, \binits{R.}}:
\bjtitle{\apj}
\bvolume{703},
\bfpage{1752}
(\byear{2009})
\end{barticle}
\endbibitem

\bibitem[\protect\citeauthoryear{{Reynolds} and {Gilmore}}{1986}]{Reynolds1986}
\begin{barticle}
\bauthor{\bsnm{{Reynolds}}, \binits{S.P.}}, \bauthor{\bsnm{{Gilmore}},
  \binits{D.M.}}:
\bjtitle{\aj}
\bvolume{92},
\bfpage{1138}
(\byear{1986})
\end{barticle}
\endbibitem

\bibitem[\protect\citeauthoryear{{Reynolds} and {Gilmore}}{1993}]{Reynolds1993}
\begin{barticle}
\bauthor{\bsnm{{Reynolds}}, \binits{S.P.}}, \bauthor{\bsnm{{Gilmore}},
  \binits{D.M.}}:
\bjtitle{\aj}
\bvolume{106},
\bfpage{272}
(\byear{1993})
\end{barticle}
\endbibitem

\bibitem[\protect\citeauthoryear{{Reynoso}}{2007}]{Reynoso2007}
\begin{barticle}
\bauthor{\bsnm{{Reynoso}}, \binits{E.M.}}:
\bjtitle{Highlights of Astronomy}
\bvolume{14},
\bfpage{305}
(\byear{2007})
\end{barticle}
\endbibitem

\bibitem[\protect\citeauthoryear{{Rothenflug}
  \textit{et~al.}}{2004}]{Rothenflug2004}
\begin{barticle}
\bauthor{\bsnm{{Rothenflug}}, \binits{R.}}, \bauthor{\bsnm{{Ballet}},
  \binits{J.}}, \bauthor{\bsnm{{Dubner}}, \binits{G.}},
  \bauthor{\bsnm{{Giacani}}, \binits{E.}}, \bauthor{\bsnm{{Decourchelle}},
  \binits{A.}}, \bauthor{\bsnm{{Ferrando}}, \binits{P.}}:
\bjtitle{\aap}
\bvolume{425},
\bfpage{121}
(\byear{2004})
\end{barticle}
\endbibitem

\bibitem[\protect\citeauthoryear{{Rybicki} and {Lightman}}{1991}]{rybicki}
\begin{bbook}
\bauthor{\bsnm{{Rybicki}}, \binits{G.}}, \bauthor{\bsnm{{Lightman}},
  \binits{A.}}:
\bbtitle{Radiative processes in astrophysics}.
\bpublisher{Wiley-Interscience},
\blocation{New-York}
(\byear{1991})
\end{bbook}
\endbibitem

\bibitem[\protect\citeauthoryear{{Sedov}}{1959}]{Sedov1959}
\begin{bbook}
\bauthor{\bsnm{{Sedov}}, \binits{L.I.}}:
\bbtitle{{Similarity and Dimensional Methods in Mechanics}}.
\bpublisher{Academic Press},
\blocation{New York}
(\byear{1959})
\end{bbook}
\endbibitem

\bibitem[\protect\citeauthoryear{{Strom}}{1988}]{Strom}
\begin{barticle}
\bauthor{\bsnm{{Strom}}, \binits{R.G.}}:
\bjtitle{\mnras}
\bvolume{230},
\bfpage{331}
(\byear{1988})
\end{barticle}
\endbibitem

\bibitem[\protect\citeauthoryear{{Takiwaki}
  \textit{et~al.}}{2004}]{Takiwaki2004}
\begin{barticle}
\bauthor{\bsnm{{Takiwaki}}, \binits{T.}}, \bauthor{\bsnm{{Kotake}},
  \binits{K.}}, \bauthor{\bsnm{{Nagataki}}, \binits{S.}},
  \bauthor{\bsnm{{Sato}}, \binits{K.}}:
\bjtitle{\apj}
\bvolume{616},
\bfpage{1086}
(\byear{2004}).
doi:\doiurl{10.1086/424993}
\end{barticle}
\endbibitem

\bibitem[\protect\citeauthoryear{{Taylor}}{1950}]{Taylor1950a}
\begin{barticle}
\bauthor{\bsnm{{Taylor}}, \binits{G.}}:
\bjtitle{Royal Society of London Proceedings Series A}
\bvolume{201},
\bfpage{159}
(\byear{1950})
\end{barticle}
\endbibitem

\bibitem[\protect\citeauthoryear{{Tziamtzis}
  \textit{et~al.}}{2011}]{Tziamtzis2011}
\begin{barticle}
\bauthor{\bsnm{{Tziamtzis}}, \binits{A.}}, \bauthor{\bsnm{{Lundqvist}},
  \binits{P.}}, \bauthor{\bsnm{{Gr{\"o}ningsson}}, \binits{P.}},
  \bauthor{\bsnm{{Nasoudi-Shoar}}, \binits{S.}}:
\bjtitle{\aap}
\bvolume{527},
\bfpage{35}
(\byear{2011})
\end{barticle}
\endbibitem

\bibitem[\protect\citeauthoryear{{Vink}}{2005}]{Vink2005}
\begin{botherref}
\oauthor{\bsnm{{Vink}}, \binits{J.}}:
In: {R.~Smith} (ed.)
X-ray Diagnostics of Astrophysical Plasmas: Theory, Experiment, and
  Observation.
American Institute of Physics Conference Series
vol. 774,
p. 241
(2005)
\end{botherref}
\endbibitem

\bibitem[\protect\citeauthoryear{{Walder} \textit{et~al.}}{2005}]{Walder2005}
\begin{barticle}
\bauthor{\bsnm{{Walder}}, \binits{R.}}, \bauthor{\bsnm{{Burrows}},
  \binits{A.}}, \bauthor{\bsnm{{Ott}}, \binits{C.D.}}, \bauthor{\bsnm{{Livne}},
  \binits{E.}}, \bauthor{\bsnm{{Lichtenstadt}}, \binits{I.}},
  \bauthor{\bsnm{{Jarrah}}, \binits{M.}}:
\bjtitle{\apj}
\bvolume{626},
\bfpage{317}
(\byear{2005}).
doi:\doiurl{10.1086/429816}
\end{barticle}
\endbibitem

\bibitem[\protect\citeauthoryear{{Yamaguchi}
  \textit{et~al.}}{2008}]{Yamaguchi2008}
\begin{barticle}
\bauthor{\bsnm{{Yamaguchi}}, \binits{H.}}, \bauthor{\bsnm{{Koyama}},
  \binits{K.}}, \bauthor{\bsnm{{Katsuda}}, \binits{S.}},
  \bauthor{\bsnm{{Nakajima}}, \binits{H.}}, \bauthor{\bsnm{{Hughes}},
  \binits{J.P.}}, \bauthor{\bsnm{{Bamba}}, \binits{A.}},
  \bauthor{\bsnm{{Hiraga}}, \binits{J.S.}}, \bauthor{\bsnm{{Mori}},
  \binits{K.}}, \bauthor{\bsnm{{Ozaki}}, \binits{M.}}, \bauthor{\bsnm{{Tsuru}},
  \binits{T.G.}}:
\bjtitle{\pasj}
\bvolume{60},
\bfpage{141}
(\byear{2008})
\end{barticle}
\endbibitem

\bibitem[\protect\citeauthoryear{{Zaninetti}}{2004}]{Zaninetti2004}
\begin{barticle}
\bauthor{\bsnm{{Zaninetti}}, \binits{L.}}:
\bjtitle{\pasj}
\bvolume{56},
\bfpage{1067}
(\byear{2004})
\end{barticle}
\endbibitem

\bibitem[\protect\citeauthoryear{{Zaninetti}}{2009}]{Zaninetti2009a}
\begin{barticle}
\bauthor{\bsnm{{Zaninetti}}, \binits{L.}}:
\bjtitle{\mnras}
\bvolume{395},
\bfpage{667}
(\byear{2009})
\end{barticle}
\endbibitem

\bibitem[\protect\citeauthoryear{{Zaninetti}}{2011}]{Zaninetti2011a}
\begin{barticle}
\bauthor{\bsnm{{Zaninetti}}, \binits{L.}}:
\bjtitle{\apss}
\bvolume{333},
\bfpage{99}
(\byear{2011})
\end{barticle}
\endbibitem

\end{thebibliography}

\end{document}